\documentclass[aps,prb,twocolumn,showpacs]{revtex4}
\usepackage{graphicx,epsf}
\usepackage{amsmath}
\usepackage{amssymb}

\newcommand{\lmag}{\xi_{\rm mag}}
\newcommand{\lconf}{\xi_{\rm conf}}
\newcommand{\lcm}{\xi_{\rm cm}}     
\newcommand{\lmax}{\xi_{\rm max}}   
\newcommand{\imp}{i_{\rm imp}}
\newcommand{\br}{{\bf r}}

\newcommand{\w}{{\omega}}

\begin{document}

\title{
Quantum magnets with weakly confined spinons:\\
Multiple length scales and quantum impurities
}
\author{R. L. Doretto}
\author{Matthias Vojta}
\affiliation{Institut f\"ur Theoretische Physik, Universit\"at zu
K\"oln, Z\"ulpicher Stra{\ss}e 77, 50937 K\"oln, Germany}
\date{\today}

\begin{abstract}
In magnets with strong quantum fluctuations,
paramagnetic ground states with or without confinement of
spinon excitations can be realized.
Here we discuss the physics of the confined phase in the vicinity
of a confinement--deconfinement transition:
this regime, likely relevant to a multitude of
frustrated spin systems, is characterized by multiple length scales.
In addition to the magnetic correlation length, a confinement
length can be defined, which can be probed, e.g., by local static
measurements near non-magnetic impurities.
We illustrate the ideas by explicit calculations for dimerized spin
chains, but our qualitative results remain valid in higher dimensions
as well.
In particular, we study crossover from weak to strong confinement
visible in the antiferromagnetic polarization cloud around
a non-magnetic impurity.
We also discuss the effective magnetic interaction between impurities,
relevant for impurity-induced magnetic order, and consequences for
nuclear magnetic resonance and neutron scattering experiments.
\end{abstract}

\maketitle

%
%


\section{Introduction}

The last decades have seen tremendous interest in exotic phases of
correlated quantum matter characterized by elementary
excitations with fractional quantum numbers. Some examples are
Luttinger liquids,\cite{giamarchi} spin liquids,\cite{lee}
and fractional quantum Hall systems.\cite{heinonen}
While it has long been known that fractionalization generically
occurs in one space dimension (1D), the situation in higher dimensions $D$ is
quite different.
It has become clear that fractionalized phases in $D\geq 2$ can be connected
to deconfined phases of certain lattice gauge theories.\cite{ss04,ss08}

Quantum antiferromagnets have played a central role in investigations of exotic
phases.\cite{ss08}
Such systems can exhibit both antiferromagnetic and paramagnetic
ground states. In the latter situation, which is generic to 1D due to
the Mermin-Wagner theorem,
quantum fluctuations prevent long-range order of the spins.
In $D\geq 2$, quantum paramagnetic states may obtained by the
formation of singlet bonds of spins 1/2.
For lattices with one spin 1/2 per unit cell,
the singlet bonds can form either crystalline order accompanied by lattice symmetry breaking,
leading to so-called valence bond solid (VBS) phases,
or short-range ordered liquid states without any broken symmetry.\cite{ss08}
Remarkably, such spin liquids can be shown to be fractionalized,
with elementary excitations being spin-1/2 spinons.

The search for microscopic realizations of spin liquids in $D \geq 2$
remains difficult, because the relevant spin models display
geometric frustration: this reduces the applicability of
otherwise powerful Quantum Monte Carlo techniques
due to the so-called sign problem.
The first unambiguous demonstration of a stable phase
with fractionalization in two-dimensional (2D) magnet was
for the triangular-lattice quantum dimer model,
which features both VBS and gapped spin liquid phases,
the latter displaying topological order.\cite{moess01}
Models with exact gapped spin-liquid ground states have been constructed,
like the toric code model\cite{kitaev03} and the spin-1/2
honeycomb model,\cite{kitaev06} both due to Kitaev.
Heisenberg models with longer-range frustrated interactions
are believed to be good candidates for topological
spin liquids as well, but with the exception of the triangular-lattice
ring-exchange model\cite{triang_ring} no conclusive numerical results are
available.
(Gapless spin liquids with power-law correlation functions do not play a role in the
subsequent discussion.)

A special form of fractionalization has been argued to be possible
at a continuous quantum phase transition between two conventional phases with
different broken symmetries.\cite{deconf}
A candidate for this scenario of deconfined quantum criticality
is the N\'{e}el--VBS transition of the square-lattice
Heisenberg antiferromagnet supplemented by a ring-exchange
interaction.\cite{sandvik}
An interesting prediction of the deconfined critical theory
is the existence of two distinct small energy scales
on the VBS side of the quantum critical point:
primarily, the transition is between a N\'{e}el state and a spin liquid,
and the VBS is a secondary instability of the spin liquid.
Consequently, dimerization and confinement set in at a much smaller
energy scale compared to paramagnetic behavior.

A common feature of (i) a confined paramagnet near a confinement--deconfinement
transition and (ii) a VBS phase near a deconfined critical point
is that, in addition to the magnetic correlation length $\lmag$,
there exists a much larger length scale associated with confinement,
which we dub confinement length $\lconf$.
In the following, we shall understand $\lconf$ as the typical spatial extension of
a two-spinon bound state.
Importantly, both $\lconf$ and $\lmag$ can be large compared to microscopic lengths,
opening the possibility to find universal phenomena characterized by two length
scales, in contrast to the conventional situation with $\lmag$ being the only large
length.
Indeed, a recent numerical study \cite{poilblanc1} of the frustrated regime
of the $J_1$--$J_2$--$J_3$ Heisenberg antiferromagnet
on a square lattice gave evidence for two distinct
length scales, with $\lconf > \lmag$.

The purpose of this paper is a detailed discussion of physical properties of quantum
paramagnets near deconfinement. We believe this regime to be relevant for a rather broad
class of frustrated magnets: Even if the elementary excitations are ultimately confined,
there will be signatures of deconfinement on shorter length scales, which can be probed
experimentally. The presence of two length scales, $\lconf$ and $\lmag$, strongly modifies a
number of phenomena, both in statics and dynamics.
We present general arguments and support them by explicit calculations
for 1D Heisenberg models.
Although the mechanisms of fractionalization are quite different
in 1D and higher dimensions, we argue that various consequences for observables
are similar, and hence our results remain qualitatively valid in
$D \geq 2$ as well.

\subsection{Observables}

In the simplest picture, weak confinement can be understood in terms of a (weak)
linear potential, $V(x) = \alpha|x|$, describing the effective interaction between two spinons.
The two-particle problem then has only bound-state solutions, i.e., all excitations are
of two-spinon type.
The lowest two-spinon triplet state takes the role of the conventional spin-1 (triplon)
excitation, its size defining the confinement length $\lconf$,
with $\lconf\to\infty$ for $\alpha\to 0$.

A static measurement of $\lconf$ is possible, e.g., by spatial pinning of one of the two partners of the
bound state.
This can be achieved by doping with non-magnetic impurities: removing one spin-1/2 from the system
breaks a singlet and liberates a spinon which now interacts with the vacancy
(holon).
Provided that the binding between spinon and holon is identical to that between two
spinons, the size of the spinon--holon bound state gives a measure of $\lconf$.
As we show below, the size of the antiferromagnetic polarization cloud around
the vacancy is given by $\lconf$ in the weakly confined regime (whereas
it is determined by $\lmag$ in the conventional case of strong confinement).
This polarization cloud, which can be detected e.g. by local magnetic probes,
also determines the effective magnetic interaction between impurity moments
and is hence relevant for impurity-induced magnetic order.

Weak confinement shows up as well in the dynamical properties, as it implies the existence
of multiple bound states below the two- or three-triplon continuum. Hence, inelastic
neutron scattering experiments, measuring the dynamic spin susceptibility, will detect a
series of sharp dispersing modes below the threshold of the multi-particle continuum.

With increasing confinement potential, $\lconf$ decreases. The criterion for the
crossover to strong confinement may be defined as $\lconf \sim \lmag$. This is
expected to coincide with the condition that the energy spacing between the lowest
two-spinon bound states becomes comparable to the spin gap, and hence the higher bound
states are no longer discernible in the dynamical susceptibility.

\subsection{Outline}

The remainder of the paper is organized as follows.
In Secs.~\ref{j1j2}--\ref{sec:exp} we discuss the physics of
weak confinement in one dimension:
Sec.~\ref{j1j2} summarizes the well-known properties of the frustrated
dimerized spin chain, described by the $J_1$--$J_2$--$\delta$ model,
and the methods applied in the following sections.
Numerical results for the undoped spin chain are discussed in
Sec.~\ref{sec:dmrg1} while analytical and numerical calculations for
the system doped with a single vacancy are presented in Sec.~\ref{sec:1imp}.
Sec.~\ref{sec:exp} describes experimental implications of weak confinement
for nuclear magnetic resonance (NMR) and neutron scattering measurements
and for impurity-induced magnetic order.
Finally, in Sec.~\ref{sec:2d} we discuss the relevance of our findings for
higher dimensions.
A brief summary of our results closes the paper.


\section{Frustrated dimerized spin chains}
\label{j1j2}

In one space dimension, spinon excitations are quite generically
deconfined. The standard spin-1/2 Heisenberg chain has a
gapless excitation spectrum, while frustration can induce
spontaneous dimerization leading to gapped (but still deconfined)
spinons. To induce confinement, one has to resort to models with
explicitly broken translational symmetry.

\subsection{Model and phase diagram}

\begin{figure}[t]
\centerline{
\includegraphics[clip,width=3.4in]{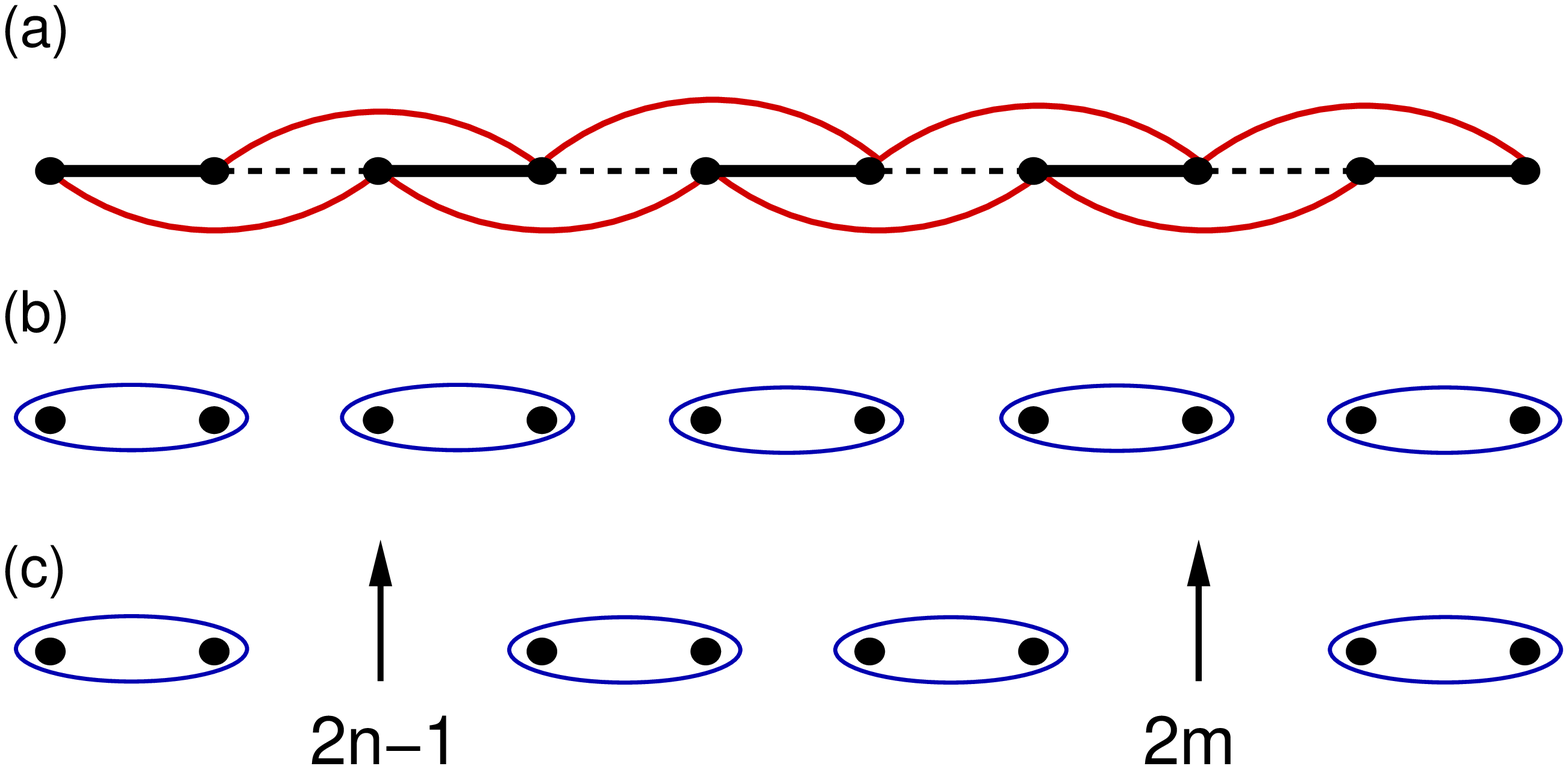}}
\caption{(color online)
Schematic representations: (a) Model
  Hamiltonian \eqref{hchain}. The thick solid (black) lines correspond to a
  strong bond [$J_1(1+\delta)$], the dashed lines to a weak bond
  [$J_1(1-\delta)$], and the thin solid (red) lines to the $J_2$
  coupling. (b) State $|0\rangle$. The ellipses represent
  singlet states between neighboring sites. (c) State $|n\,m\rangle$.
  The two spins at sites $2n-1$ and $2m$ are separated by a
  singlet pattern.
}
\label{ham-spinons}
\end{figure}

We consider the so-called $J_1$--$J_2$--$\delta$ model,
describing a frustrated dimerized antiferromagnetic (AF) spin chain:
\begin{eqnarray}
H_{ch} &=& J_1\sum_{i=1}^{2N}\left(1 - (-1)^i\delta\right){\bf S}_i\cdot{\bf
  S}_{i+1} +  J_2\sum_{i=1}^{2N}{\bf S}_i\cdot{\bf S}_{i+2}.
\nonumber \\
\label{hchain}
\end{eqnarray}
Here ${\bf S}_i$ represents a spin-1/2 at site $i$,
$J_1$ and $J_2$ are the antiferromagnetic first- and second-neighbor
exchange couplings, and $\delta$ denotes the
degree of dimerization, see Fig.~\ref{ham-spinons}(a).
(We will employ $J_1=1$ in the following.)

\begin{figure}[t]
\centerline{
\includegraphics[clip,width=2.6in]{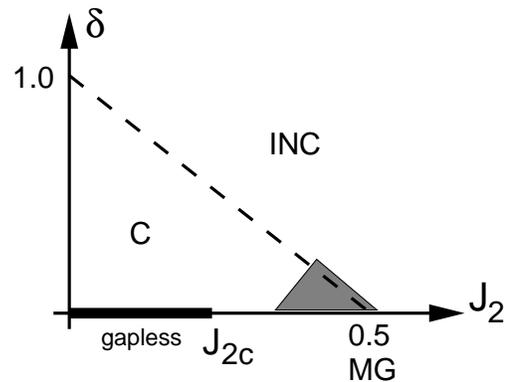}}
\caption{
Schematic ground-state phase diagram of the frustrated dimerized
Heisenberg model \eqref{hchain}.
For $\delta=0$, the system is gapless for $J_2 < J_{2C} = 0.2411$
and spontaneously dimerized for $J_2 > J_{2C}$,
whereas $\delta>0$ induces a gap for all $J_2$.
On the so-called disorder line, $2J_2 + \delta = 1$ (dashed),
the ground state is given by a singlet product state on the strong bonds;
moreover, the disorder line separates regimes of commensurate (C) and
incommensurate (INC) spin correlations.
The shaded area is the weakly confined regime of main interest in this paper,
with $\lconf\gg\lmag$.
}
\label{fig:pd_chain}
\end{figure}

The ground-state phase diagram of the model \eqref{hchain} in the $J_2$--$\delta$
plane is well known,\cite{chitra,swapan} Fig.~\ref{fig:pd_chain},
and will be summarized in the following.
We start with $\delta = 0$.
For small $J_2 < J_{2C}$, the ground state is unique and the
excitation spectrum is gapless.
At the critical value $J_{2C} = 0.2411$,\cite{okamoto,affleck97}
a continuous phase transition of Kosterlitz-Thouless type into a gapped phase takes place.
Here, the system breaks an Ising symmetry accompanied by bond order, i.e., spontaneous
dimerization.
At $J_2 = 0.5$, the so-called Majumdar-Ghosh (MG) point, the two ground-state
wavefunctions are given by a nearest-neighbor singlet product states:\cite{majumdar}
\begin{eqnarray}
|0\rangle &=& [1,2][3,4][5,6]\ldots [2N-1,2N],
\nonumber \\
\label{singlets} && \\
|\bar{0}\rangle &=& [2N,1][2,3][4,5]\ldots [2N-2,2N-1],
\nonumber
\end{eqnarray}
with $[i,j]$ denoting a normalized singlet combination of the spins
at sites $i$ and $j$,
\begin{equation}
[i,j] = \frac{1}{\sqrt{2}}\left[|\uparrow\rangle_i|\downarrow\rangle_j
  -                            |\downarrow\rangle_i|\uparrow\rangle_j\right].
\label{singlet-state}
\end{equation}
The state $|0\rangle$ is illustrated in Fig.~\ref{ham-spinons}(b).

In the gapped phase, elementary excitations are
topological defects (solitons or spinons) separating the states
$|0\rangle$ and $|\bar{0}\rangle$. Shastry and
Sutherland\cite{shastry} performed variational calculations for the MG
point using a
reduced Hilbert space $\{|n\, m\rangle\}$, where the states $|n\,
m\rangle$ consist of two spins at sites $2n-1$ and $2m$ separated by
nearest-neighbor singlets, Fig.~\ref{ham-spinons}(c).
It was shown that the
low-lying excitations are pairs of propagating spinons, i.e., deconfined
spin-1/2 (fractionalized) quasiparticles.
The upper bound for the spinon dispersion, $\epsilon_k = (1/2)(5/4
+ \cos2k)$, was later confirmed by numerical calculations.\cite{sorensen}

A finite dimerization $\delta$ opens a gap for all $J_2$.
The perfectly dimerized states $|0\rangle$ and $|\bar{0}\rangle$
remain eigenstates of $H_{ch}$ along the so-called ``disorder line''
$2J_2 + \delta = 1$, but their degeneracy is lifted by $\delta>0$.
The energy of the ground state $|0\rangle$ is $E_0 = (-3/8)(1+\delta)(2N)$.
It was also shown that the static structure factor
$S(q) = (1/2N)\sum_j\exp(iqR_j)\langle {\bf S}_0\cdot{\bf S}_j\rangle$
peaks at momentum $q = \pi$ and $q < \pi$, respectively, for
points localized on the left (commensurate phase) and right
(incommensurate phase) sides of the disorder line in the phase
diagram.\cite{chitra}

The fact that the states $|0\rangle$ and $|\bar{0}\rangle$ are no
longer degenerate implies that the low-lying excitations are now
bound states of two spinons, i.e., spin-1
confined objects.\cite{sorensen}
Indeed, a finite value of $\delta$ introduces an
effective attractive potential between two spinons: by converting one
particular singlet of the state  $|0\rangle$ into a triplet and
then separating the two spinons as in the state $|n\, m\rangle$, the energy
of the system increases by an amount proportional to
$|m-n|\delta$. Such an increase is due
to the fact the singlets between the two spinons are now
located at the weak bonds [compare Figs.~\ref{ham-spinons}.(a) and
(c)].\cite{uhrig}
As noted above, the spatial extension of the lowest-energy spinon--spinon
bound state\cite{BSfoot} provides a measure of the confinement length $\lconf$,
with $\lconf\to\infty$ for $\delta\to 0$.

In the following, we shall be interested in the spin-gapped regime of weak confinement,
with $\lconf\gg\lmag$. Thus we shall work at $\delta \ll 1$ and $J_2$ close to 1/2,
Fig.~\ref{fig:pd_chain}.
As the data analysis is much simpler for commensurate (instead of incommensurate)
spin correlations, the regime to the right of the disorder line will be touched upon only
briefly.

\subsection{Methods}

Ground-state properties of the spin chain \eqref{hchain} will be calculated
numerically for finite-size systems with the aid of the density-matrix renormalization
group (DMRG) method,\cite{dmrg} using the libraries provided by the ALPS collaboration.\cite{alps}
The calculations for $2N = 100\ldots400$ sites employ open boundary conditions,
such that strong bonds are located at the chain ends, Fig.~\ref{ham-spinons} (a).

In addition, near the disorder line, variational wavefunctions consisting of
nearest-neighbor singlets with additional spinons will be utilized, similar to those
in Ref.~\onlinecite{shastry}. Details can be found in
Appendix \ref{ap:variational}.


\section{Spin correlations of undoped spin chains}
\label{sec:dmrg1}

As a reference, we calculate the bulk spin correlation function
$\langle S_i^Z S_j^Z\rangle$ in the ground state of the Hamiltonian
\eqref{hchain} using DMRG.
We restrict our attention to small $\delta$ and the vicinity of the
disorder line.

\begin{figure}[t]
\centerline{
\includegraphics[clip,width=2.7in]{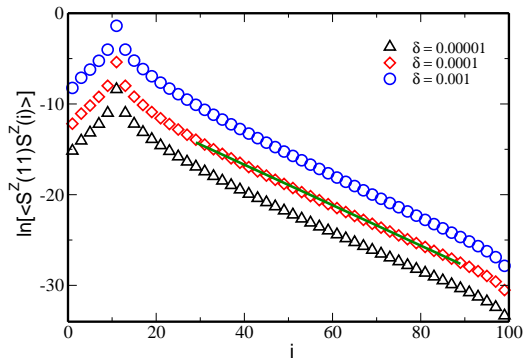}}
\caption{(color online)
  Natural logarithm of the spin-spin correlation function
  $\langle S^Z_{i=11}S^Z_i \rangle$ for $J_2 = 0.45$ and $\delta =
  10^{-5}$, $10^{-4}$, and
  $10^{-3}$. The results are from DMRG calculations with $m = 128$
  states for the spin chain \eqref{hchain} with $2N = 100$ sites. The
  curves are vertically shifted for visualization. Only the sites
  with $\langle S^Z_{i=11}S^Z_i \rangle > 0$ are shown. The solid
  (green) line is an exponential fit to the data.
}
\label{log-sz-sz}
\end{figure}

The typical behavior of $\langle S^Z_iS^Z_j \rangle$
within the bulk of the chain for $J_2 =
0.45$ and several values of $\delta$ is shown in
Fig.~\ref{log-sz-sz} on a log-linear plot. One can clearly see an
exponential decay of the spin correlations, i.e.,
$\langle S^Z_iS^Z_j \rangle \sim \exp\left(|i-j|/\lmag\right)$,
as it is expected for a spin gapped system.
Similar results were found for $J_2 = 0.48$.

\begin{figure}[t]
\centerline{\includegraphics[clip,width=2.7in]{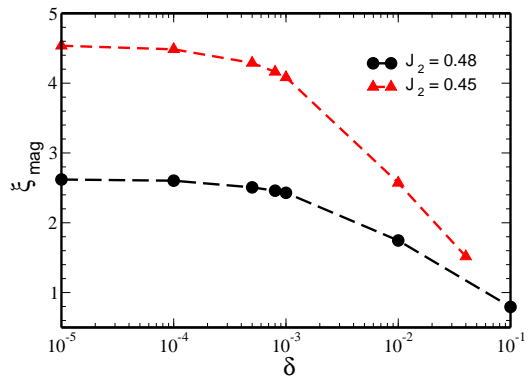}}
\caption{(color online)
The magnetic correlation length for $J_2 = 0.45$ and $J_2 =
  0.48$ in terms of $\delta$ calculated via DMRG (see text for the
  details of the calculation). The dashed lines are guide for the eyes.}
\label{lmag}
\end{figure}

In Fig.~\ref{lmag}, we show the values of $\lmag$, obtained from
an exponential fit to the long-distance part to the data, as a
function of $\delta$ for $J_2 = 0.45$ and $0.48$. (In the following,
all length scales are measured in units of the lattice spacing.)
For a fixed value of $J_2$, $\lmag$ increases as the dimerization
decreases and then approaches a constant value in the limit $\delta
\rightarrow 0$. $\lmag$ increases as $J_2$ is reduced for a
fixed value of $\delta$. This behavior is in agreement with the
discussions of Sec.~\ref{j1j2}. Recall that along the line $\delta =
0$, $\lmag$ diverges as $J_2 \to J_{2C}$ from above.
We note that {\em on} the disorder line $\lmag$ is minimal,
but the correlations cannot be reasonably fitted to an exponential decay,
as the ground state of the infinite system is given by the product state
$|0\rangle$ \eqref{singlets}.

\section{Spin chains with a non-magnetic impurity}
\label{sec:1imp}

Vacancy doping in quantum paramagnets can induce effective magnetic moments.
These impurity-induced moments reveal themselves in a Curie-like behavior of the
uniform susceptibility, $\chi \propto C/T$, at intermediate temperatures.
Remarkably, in the presence of three-dimensional couplings
these induced moments can order at sufficiently low temperatures,
thus changing the spin-gapped paramagnetic ground state of the pure compound
into a magnetically long-range ordered state upon doping.

The appearance of effective moments upon doping vacancies
is a property of systems with confined spinons;
for spin $1/2$ systems it is best visualized in terms of broken singlet bonds
where one spin is replaced by a vacancy.
The liberated spin $1/2$ is confined to the vacancy
at low energies, resulting in an effective spin-$1/2$ moment.\cite{sigrist,icmp}
In contrast, in host systems with deconfined spinons, no moments are generated
by introducing vacancies.
This theoretical picture has been supported by various numerical
studies, in particular on spin chain\cite{bruce} and ladder
systems.\cite{elbio,Miyazaki,Yasuda}

In the following, we present analytical and DMRG results for spin chains with a single vacancy,
with focus on the spin polarization cloud associated with the impurity-induced magnetic
moment.
It has been established that, in the strongly confined regime, the cloud consists
of excited triplets, with the spatial size given by $\lmag$.\cite{elbio,Miyazaki,Yasuda,vbs,ssmv}
In contrast, in the weakly confined regime the cloud size is given by the
extension of the spinon--holon bound state, i.e., $\lconf$.
Indeed, we shall find a cloud of size ${\rm max}(\lmag,\lconf)$,
with a distinct crossover for $\lmag \sim \lconf$.
The shape of the spin polarization cloud allows us to distinguish
the weakly from the strongly confined regime.

A remark on limits is in order here: the ground state of the infinite system with a
single vacancy is doubly degenerate, corresponding to the residual vacancy-induced
moment. Applying a uniform field $h_u \ll \Delta$, where $\Delta$ is the bulk spin gap,
fully polarizes this moment and leads to a static staggered spin arrangement near the vacancy.
A finite concentration of vacancies introduces another energy scale, namely the effective
(maximum) interimpurity interaction $J_{\rm eff}$, see also Sec.~\ref{sec:imp-int}.
The following discussion of the polarization cloud around a vacancy assumes the limit
$h_u\to 0$, keeping $J_{\rm eff} \ll h_u \ll \Delta$.


\subsection{Variational calculations}
\label{sec:var}

\begin{figure}[t]
\centerline{\includegraphics[clip,width=3.4in]{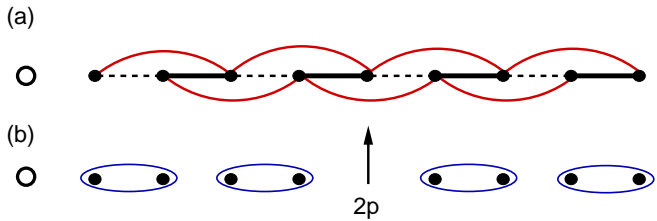}}
\caption{(color online)
Schematic representation of (a) the Hamiltonian \eqref{hchain}
  with a non-magnetic impurity (open circle) at site
  $i=1$ and (b) the state $|p\rangle$ which consists of an empty site
  at $i=1$ and a spinon at site $2p$ separated by a
  singlet pattern. The symbols are the same as in Fig.~\ref{ham-spinons}.
}
\label{imp-ham}
\end{figure}

Near the disorder line
the low-energy states of the dimerized spin chain \eqref{hchain} are dominated
by short-range singlet coverings:
The $S=1$ states $|m\,n\rangle$, Fig.~\ref{ham-spinons}(c), have a lower energy
than any $S=2$ state even for a large separation $(m-n)$ between the two spinons.
This is the basis of the variational method of Ref.~\onlinecite{shastry},
which we adopt here to calculate the spin correlations in the vicinity of a
non-magnetic impurity.

Consider the frustrated dimerized spin chain \eqref{hchain} with
strong bonds at the chain ends and introduce a vacancy at an odd site.
For finite dimerization, $\delta>0$, the system has no inversion symmetry at
the impurity site $\imp$ even in the infinite-system limit.
In this configuration, the liberated spin-1/2 stays
mainly in the region $ i > \imp$:
recall that the low-energy sector is formed by states made out of short-range
singlets. The free spin can only  occupy a site $ i < \imp$
if a (long) singlet is formed between the spins at sites $\imp -1$ and $\imp+ 1$,
which corresponds to a high-energy state.
To simplify the variational calculations, we shall therefore ignore the region $ i < \imp$,
i.e., consider a spin chain with the vacancy placed at the chain end,
as illustrated in Fig.~\ref{imp-ham}(a).
Also, we focus on parameter values on the disorder line,
$J_2 = (1 - \delta)/2$ with $10^{-5} \le \delta \le 10^{-2}$.

We restrict the variational Hilbert space to
the states $|p\rangle$, which consist of an empty site at $i=1$ and a
spinon at site $2p$ in a background of nearest-neighbor singlets as
illustrated in Fig.~\ref{imp-ham}(b).
The states $|p\rangle$ are not orthogonal to each other. In fact, the
overlap between two states $|m\rangle$ and $|p\rangle$ reads
\begin{equation}
\langle m | p \rangle = \left(-\frac{1}{2}\right)^{|m - p|}.
\label{overlap}
\end{equation}
A long but straightforward calculation shows
that the matrix elements of the Hamiltonian \eqref{hchain} with a
vacancy at site $i=1$ are given by
\begin{eqnarray}
&& 2 \langle m |H - E_0| p \rangle = \frac 1 2 \big\{ 2 + 5J_2
   \nonumber \\
\nonumber && \\
  &&+ \left. 3(1-2J_2) \left[ 1 + \theta_+(m,p)  +
      (m-p)\theta_-(m,p)\right]\right. \nonumber \\
\nonumber && \\
  &&+ \left.  3\delta\left[2p - 1 + \theta_-(m,p) \right.\right.
    + \left. (m-p)\theta_+(m,p)\right]\big\}\langle m | p
  \rangle  \nonumber \\
\nonumber && \\
 &&+ (1 + \delta - J_2)\langle m | p - 1\rangle
   + (1 - \delta - J_2)\langle m | p + 1\rangle,
\label{matrix}
\end{eqnarray}
where $E_0 = -(3/8)(1 + \delta)(2N)$, $\theta_+(m,p) \equiv
\theta(p-m-1) + \theta(m-p-1)$,
and $\theta_-(m,p) \equiv \theta(p-m-1) - \theta(m-p-1)$ with the step
function $\theta(x)$ defined as in Eq.~\eqref{theta}.

Instead of considering the non-orthogonal basis $\{|p\rangle\}$, we can
define a orthonormal one\cite{uhrig} via the Gram-Schmidt orthogonalization
procedure, namely
\begin{eqnarray}
\nonumber
|\phi_1\rangle &\equiv& |1\rangle, \\
|\phi_p\rangle &=& \frac{2}{\sqrt{3}}\left[|p\rangle + (1/2)|p-1\rangle\right],
\;\;\;\;\;\;\;\; p\ge 2.
\label{basis}
\end{eqnarray}
Diagonalizing the Hamiltonian \eqref{imp-ham} in the basis
$\{|\phi_p\rangle\}$ gives the ground-state energy $E$ and
wavefunction $|\Psi\rangle$ in the presence of an impurity:
\begin{equation}
|\Psi\rangle = \sum_{p=1}^NC_p|\phi_p\rangle.
\label{ansatz}
\end{equation}

\begin{figure}[t]
\centerline{\includegraphics[clip,width=2.7in]{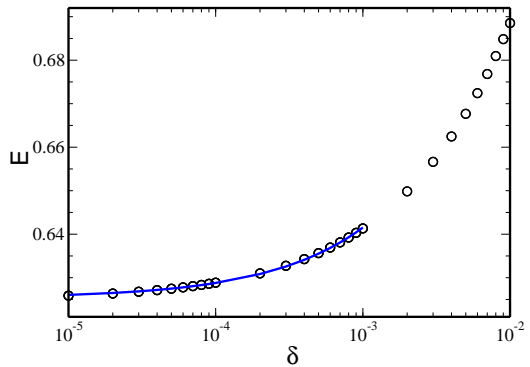}}
\caption{(color online)
  Variational ground state energy $E$ as a
  function of $\delta$ (open circles)
  for the doped frustrated dimerized spin chain. The
  results are for
  points along the disorder line, i.e., $J_2 = (1 - \delta)/2$. The
  solid (blue) line is a power law fit to the data in the region $10^{-5} \le
  \delta \le 10^{-3}$ (see text for details).}
\label{energy-var}
\end{figure}

\begin{figure}[t]
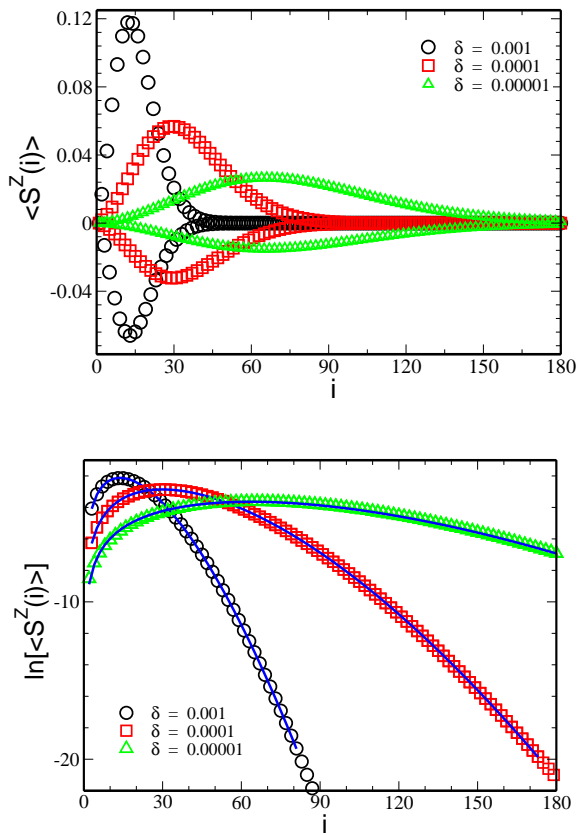

\centerline{
\includegraphics[clip,height=5.2cm]{magnetizacao104-v2-linear.eps}}
\vskip0.75cm
\centerline{
\includegraphics[clip,height=5.0cm]{magnetizacao104-v3.eps}}
\caption{(color online)
Variational results for the spin
polarization $\langle S^Z_i \rangle$ around $\imp = 1$. $J_2 = (1 - \delta)/2$ and
$\delta=10^{-3}$ (black circles), $10^{-4}$ (red
squares), and $\delta=10^{-5}$ (green triangles).
Top panel: linear-linear scale.
Bottom panel: same data on a log-linear scale, showing only the sites
with $\langle S^Z_i \rangle > 0$; the solid (blue) lines are fits to
the data using Eq.~\eqref{fit} for the entire data set with $i>1$
and $\ln[\langle S^Z_i \rangle] > -20$.
}
\label{sz-var}
\end{figure}

In Fig.~\ref{energy-var}, we show the energy $E$ as a function of
$\delta$ calculated for a spin chain with $2N = 600$ sites using the
procedure described above.
The data in the region $10^{-5} \le \delta \le 10^{-3}$, which, as we
will see in the next section, is deep in the weakly confined regime, can be fitted
by a power law $a_0 + a_1\delta^{2/3}$ with $a_0 = 0.6252$ and $a_1 = 1.6241$.
Such $\delta$ dependence of the ground state energy is chosen based on
the analytical findings of Uhrig and collaborators,\cite{uhrig} who studied a similar
problem via an effective one-dimensional Schr\"odinger equation. They
showed that the solutions $\psi(x)$ of the effective equation can be
written as $\psi(x) \sim {\rm Ai}(x/\xi + z_i)$, where $z_i$ are the
zeros of the Airy function ${\rm Ai}(x)$\cite{handbook} and $\xi =
(3m\delta/2)^{-1/3}$ is the characteristic length scale with $m$ being
the spinon mass, and the ground state energy $E \sim \delta^{2/3}$.

Once the coefficients $C_p$ are known, the spin polarization
$\langle S^Z_i \rangle = \langle \Psi | S^Z_i|\Psi\rangle =
\sum_{p,m}C^*_mC_p\langle \phi_m | S^Z_i |\phi_p \rangle$
can be calculated with the aid of Eq.~\eqref{basis} and the expression
\begin{eqnarray}
\langle m | S^Z_i |p \rangle &=& (-1)^i\frac{1}{2}\left[
  \theta(i-2m)\theta(2p-i) +\theta(i-2p) \right. \nonumber \\
\nonumber && \\
  && \left. \times\theta(2m-i)\right]
  \left(1 - (1/2)\delta_{m,p}\right)\langle m | p \rangle.
\label{matrix-sz}
\end{eqnarray}
The results for $\delta=10^{-5} \ldots 10^{-3}$ are shown in
Fig.~\ref{sz-var} on a linear-linear and log-linear scales.
One can see that the spin polarization has a staggered profile in the
vicinity of $\imp$ as is was already  realized in previous
calculations.\cite{sorensen,elbio,fukuyama}
By decreasing the dimerization, not only the
position of the maximum of $\langle S^Z_i \rangle$ moves away from the
impurity site but also the distribution gets broader and broader.
Such features indicate that the system approaches
a deconfined regime as the dimerization decreases. Indeed, at $\delta
= 0$, the spin polarization peaks at the center of the chain.

\begin{figure}[t]
\centerline{
\includegraphics[clip,width=2.7in]{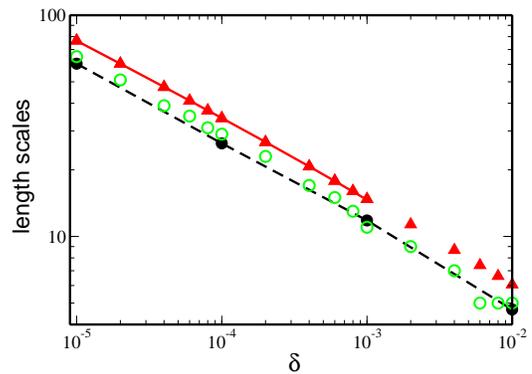}}
\caption{(color online)
Variational $\lcm$ (filled red triangles), $\lmax$ (open green circles) and
  $\xi_1$ (filled black circles) as a function of $\delta$ for the disorder
  line. The solid (red) line corresponds to a power law fit to the
  data in the region $10^{-5} \le \delta \le 10^{-3}$
 (see text for details) and the dashed line is a guide for the eyes.
}
\label{lconf-var}
\end{figure}

We now discuss possible quantitative measures of the size of the polarization cloud.
This is not trivial, as the long-distance decay deviates from a pure exponential, in
particular for small $\delta$.
The asymptotic behavior of the Airy function ${\rm Ai}(x)$ suggests
a fit using the expression
\begin{equation}
\log \langle S^Z_i \rangle  = b_1 - [|i-\imp|/\xi_1]^{1+b_3^2} + b_2\log|i-\imp|
\label{fit}
\end{equation}
for $i>\imp$, with parameters $b_{1\ldots3}$ and $\xi_1$, the latter being the length scale
corresponding to the cloud size.
This formula, being able to interpolate between a pure exponential and a faster decay,
depending on whether $b_3=0$ or $b_3\neq 0$, nicely fits the data as is illustrated in
Fig.~\ref{sz-var}.
Alternative definitions for the polarization cloud size (to the right of $\imp$)
are given by the position of the
maximum $\lmax$ of $\langle S^Z_i \rangle$ relative to $\imp$
and by the center-of-mass coordinate of the cloud:\cite{poilblanc03}
\begin{equation}
\lcm = \frac{\sum_{i=1}^{2N}(i-\imp)|\langle S^Z_i
  \rangle|}{\sum_{i=1}^{2N}|\langle S^Z_i \rangle|}.
\label{xiconf}
\end{equation}
The variational results for $\lcm$, $\lmax$, and $\xi_1$
as a function of $\delta$ are plotted in Fig.~\ref{lconf-var}.
As expected, the cloud size diverges as $\delta \rightarrow 0$,
which corresponds to the MG point,
and all measures are much larger than $\lmag \sim 1$.
For instance, the data for $\lcm$ in the region $10^{-5} \le \delta
\le 10^{-3}$ can be fitted by $a_0 + a_1\delta^{-1/3}$ with $a_0 =
-2.3952$ and $a_1 = 1.7031$. Again, the $\delta$ dependence follows
from the results derived in Ref.~\onlinecite{uhrig}.

\subsection{Numerical results}
\label{sec:dmrg2}

\begin{figure}[t]
\centerline{\includegraphics[clip,width=3.4in]{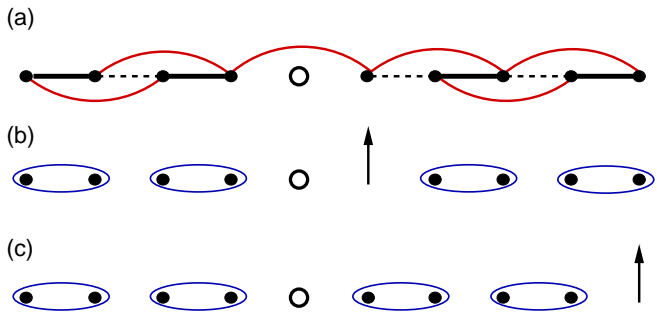}}
\caption{(color online)
Schematic representation of (a) the Hamiltonian
\eqref{hchain} with a non-magnetic impurity (open circle) at site
$\imp$ and the ground-state configurations for (b) strongly and
(c) weakly confined regimes.
The symbols are the same as in Fig.~\ref{ham-spinons}.
}
\label{imp-ham2}
\end{figure}

We now use DMRG to investigate spin chains with a single vacancy
in a parameter regime away from the disorder line.
We keep $m=150$ states for a chain with $2N = 400$ sites and open boundary condition,
and the vacancy is located at site $\imp = 101$, Fig.~\ref{imp-ham2}(a).
This choice is due to the fact that the spin polarization cloud around
$\imp$ is asymmetric (see the discussion below).
In the following, we only discuss the properties of lowest state in the
$S^Z_{tot} = -1/2$ sector.

\begin{figure}[b]
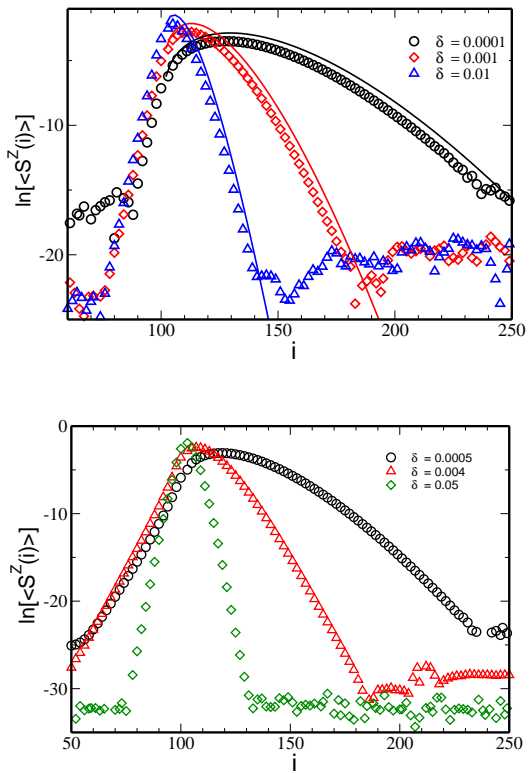

\centerline{
\includegraphics[clip,height=4.7cm]{local-sz-400-imp101-log-disorder.eps}}
\vskip0.75cm
\centerline{
\includegraphics[clip,height=4.7cm]{local-sz-400-imp101-log-v2.eps}}
\caption{(color online)
DMRG results for the spin polarization $\langle S^Z_i \rangle$
around the vacancy in the $J_1$--$J_2$--$\delta$ chain,
for parameters on the disorder line (top panel) and for
$J_2 = 0.48$ (bottom panel).
The system size is $2N = 400$, and the vacancy is at site $i=101$.
Only the sites with $\langle S^Z_i \rangle > 0$ are shown.
The solid lines in the top panel are the
variational results calculated in Sec.~\ref{sec:var}.
}
\label{sz-dmrg}
\end{figure}

Some representative results for the spin polarization around the vacancy
are plotted in Fig.~\ref{sz-dmrg} on a log-linear
scale (only $\langle S^Z_i \rangle > 0$ are shown).
For the disorder line, the spin polarization curves are asymmetric
with respect to $\imp$.
On the left side of the impurity ($i < \imp$) an
exponential tail is evident while on the right side ($i > \imp$) the
shape of the spin polarization resembles the one described in the
last section. In fact, the results of DMRG and the variational approach (solid lines in
Fig.~\ref{sz-dmrg}) are in quite good agreement.
The behavior for $J_2 = 0.48$ line is richer:
Notice that for the largest value of $\delta$, $\langle S^Z_i \rangle$  is
symmetric around the impurity site with exponential tails. As
$\delta$ decreases, the curves become more and more asymmetric, i.e.,
they display features qualitatively similar to the ones observed for
the disorder line. Indeed for $\delta < 10^{-2}$ a pure exponential decay is
no longer visible in the region $i > \imp$.
Similar behavior (not shown here) is observed for $J_2 = 0.45$.
We have also studied the vacancy-induced spin polarization for $J_2>1/2$:
in this regime, the polarization displays incommensurate oscillations with an envelope
similar to those in Fig.~\ref{sz-dmrg}. However, the incommensurate behavior complicates
numerical fits such that we refrain from a further analysis.

\begin{figure}[t]
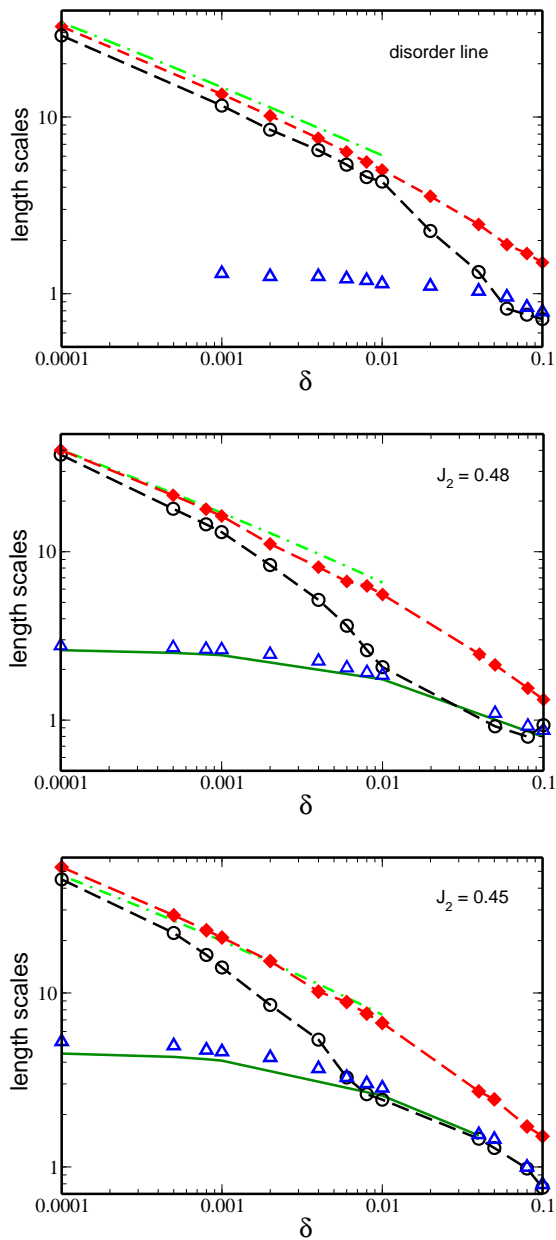

\centerline{\includegraphics[clip,height=5.1cm]{corr-length-400-disorder-v5.eps}}
\vskip0.5cm
\centerline{\includegraphics[clip,height=5.1cm]{corr-length-048-400-v5.eps}}
\vskip0.5cm
\centerline{\includegraphics[clip,height=5.1cm]{corr-length-045-400-v5.eps}}
\caption{(color online)
The characteristic length scales $\xi_1$ (open circles) and $\xi_2$
(open triangles), $\lcm$ (diamonds), and $\lmag$ (solid line)
as function of $\delta$
for $J_2=(1-\delta)/2$, 0.48, 0.45 from top to bottom,
extracted from DMRG (see text for details).
The variational $\lcm$ (dashed-dotted line) is also shown for comparison.
The dashed lines are guide to the eyes.
Note that (i) $\lmag$ cannot be meaningfully defined for $J_2=(1-\delta)/2$, i.e.,
on the disorder line, and that (ii) $J_2 = 0.48$ and
$\delta > 0.04$ corresponds to the incommensurate region.
}
\label{lconf-dmrg}
\end{figure}

We proceed as in Sec.~\ref{sec:var} and perform a quantitative
analysis of the DMRG data for $\langle S^Z_i \rangle$.
Our results are summarized in Fig.~\ref{lconf-dmrg}
which shows the decay lengths $\xi_1$, $\xi_2$ (described below),
the center-of-mass position of the cloud, $\lcm$ \eqref{xiconf},
and the bulk correlation length $\lmag$ (Sec.~\ref{sec:dmrg1})
as function of $\delta$ for three different values of $J_2$.
We also include the variational $\lcm$ determined in the previous section.
One can see that the variational and DMRG results for
$\lcm$ along the three lines are in quite good agreement.
This implies that the restriction of the Hilbert space to the subspace
$\{|p\rangle\}$ is a reasonable approximation to describe the
(weak) confinement physics. Recall that $|p\rangle$ includes only singlets
made out of neighboring sites and the ones between the
$\imp$ and the spinon site are located at the weak bonds in contrast
to the ground state of the {\it undoped} chain whose singlet pairs are placed at
the strong bonds.

The spatial decay of the spin polarization cloud, $\langle S^Z_i \rangle$,
can be analyzed with the aid of the expression \eqref{fit}.
As $\langle S^Z_i \rangle$ is asymmetric due to the broken inversion symmetry
with respect to $\imp$, we use Eq.~\eqref{fit} separately for
the regions $i>\imp$ and $i<\imp$ to extract the length scales $\xi_1$ and $\xi_2$,
respectively.
In both cases, we only include data points into the fit where $\langle S^Z_i \rangle$ is
above the noise floor of the numerical data, e.g., $\imp < i < 140$ in the $\xi_1$ fit
of the $\delta=0.01$ data in the top panel of Fig.~\ref{sz-dmrg}.
The $\xi_1$ and $\xi_2$ resulting from the fit are shown in Fig.~\ref{lconf-dmrg}
as well.

Let us discuss the results collected in Fig.~\ref{lconf-dmrg}.
In all panels, one can identify three distinct regimes:
(a) For large $\delta$, we have $\xi_1 \approx \xi_2 \approx \lmag$,
i.e., both parts of the spin polarization cloud are determined by
magnetic (bulk) correlations. We identify this as the strongly confined
regime.
(b) For small $\delta$, $\xi_1 \approx \lcm \gg \xi_2 \approx \lmag$.
Hence, the right part of the polarization cloud is determined by the
confinement length, which is moreover much larger than the bulk correlation length.
These are characteristics of the weakly confined regime, and
we conclude $\xi_1=\lconf$.
(c) There is a pronounced crossover regime between (a) and (b),
e.g. $10^{-3} \lesssim \delta \lesssim  10^{-2}$ for $J_2=0.48$.
As anticipated above, $\lconf$ diverges as $\delta\to 0$
while $\lmag$ approaches a constant value -- this proves the existence
of two length scales in the regime of weak confinement.

Qualitatively, the behavior of the spin polarization cloud is easily
understood: for strong confinement, i.e. large $\delta$, we have
$\lconf \sim 1$ which implies that the spinon is close to $\imp$.
Fig.~\ref{imp-ham2}(b) illustrates the most probable configuration of
the ground state. Notice that the singlet pairs are at the strong
bonds as in the ground state of the {\it undoped} chain. Therefore the
enhancement of the spin correlations around $\imp$ is mainly due to
triplets which are excited by the coupling between the spin adjacent
to the impurity (i.e. the spinon) and the
environment.\cite{elbio,Miyazaki,Yasuda,vbs,ssmv} The triplet physics
is dominated by $\lmag$, hence $\xi_1\approx\xi_2\approx\lmag$. On the
other hand, in a weakly confined regime of small $\delta$, $\lconf \gg
1$. The spinon travels to distant sites $i>\imp$,
Fig.~\ref{imp-ham2}(c), where it may also generate triplets. In
contrast, configurations with the spinon at sites $i<\imp$ are
suppressed as they require a long, i.e., high-energy, singlet. Hence,
the impurity-induced spin polarization for $i>\imp$ profits from the
travelling spinon and is determined by $\lconf$ (provided that
$\lconf>\lmag$) whereas the polarization for $i<\imp$ decays on the
bulk scale $\lmag$.

To summarize, the size and shape of the spin polarization induced by a
non-magnetic impurity allow to distinguish the regimes of weak and strong
confinement. In the present 1D case, both length scales $\lconf$ and $\lmag$
can in principle be extracted from $\langle S^Z_i \rangle$.


\section{Weak confinement in experiments}
\label{sec:exp}

In this section, we briefly sketch how weak confinement manifests itself in standard
experiments. As before, the discussion is focused on 1D systems, but qualitatively
applies to $D\geq2$ as well, see Sec.~\ref{sec:2d}.


\subsection{Dynamic susceptibility: Multiple bound states}
\label{dynamics1}

We start with the zero-temperature dynamic susceptibility $\chi(\vec q,\w)$
of a clean (i.e. impurity-free) gapped confined paramagnet.
The minimum of the triplon dispersion is located at some wave vector $\vec q=\vec Q$,
where $\chi(\vec q,\w=0)$ displays a maximum.

In the conventional case of strong confinement, $\chi''(\vec Q,\w)$ consists of a single
sharp peak at the spin-gap energy $\Delta$, corresponding to the triplon particle, and a
continuum of multi-particle states, which typically starts at $3\Delta$.
(In low-symmetry situations, like the $J_1$--$J_2$--$\delta$ chain,
a two-particle continuum starting at $2\Delta$ at $\vec q=\vec 2Q$ is present.)
In contrast, in a situation of weak confinement, there will be additional sharp peaks
between $\Delta$ and the onset of the continuum. This can be easily understood:
while $\Delta$ is the energy of the lowest spin-1 two-spinon bound state, weak confinement
implies that higher two-spinon bound states are energetically only slightly above $\Delta$.
The crossover to strong confinement may be defined by the criterion that $\Delta$ becomes
equal to the energetic difference between the two lowest $S=1$ two-spinon states.

For the $J_1$--$J_2$--$\delta$ chain considered above,
the dynamic structure factor has been
discussed e.g. in Ref.~\onlinecite{byrnes}. The energy spacing of the two-spinon bound states
scales as $\delta^{-2/3}$, i.e. the number of bound states below the continuum diverges
as $\delta\to 0$. The precise structure of $\chi''(\vec Q,\w)$ is determined by the
energies and matrix elements of the two-spinon bound states,
the latter giving the weights of the peaks in $\chi''$.
The energies depend on the bound state number $n$ as $\Delta+n^{2/3}$.
Ref.~\onlinecite{byrnes} also provides an estimate of the peak weights,
but the employed continuum limit is questionable because the weight depends
on the short-distance behavior of the two-particle wavefunction which
is strongly influenced by lattice effects due to the non-analytic nature
of the confinement potential $V(x) = \alpha|x|$.
Numerical results for $\chi''(\vec q,\w)$ in the $J_1$--$J_2$--$\delta$ chain
can be found in Refs.~\onlinecite{sorensen,trebst1,trebst2,uhrig04}, which clearly show
additional sharp modes below the continuum.
In particular, Fig.~4 of Ref.~\onlinecite{trebst1} reports energies of {\em triplon} bound states
on the disorder line for $\delta=0.4$.
As both single-triplon states and two-triplon spin-1 bound states correspond to
two-spinon spin-1 bound states, their results can be directly cast in our language:
there are two (four) two-spinon spin-1 bound states below the two-particle continuum
at wave vector 0 ($\pi/2$),
but already the second is very close to the lower edge of the continuum, consistent
with $\delta=0.4$ being located in the crossover regime between
weak and strong confinement, Fig.~\ref{lconf-dmrg}.


\subsection{NMR line shape due to vacancies}
\label{sec:nmr}

We now return to confined paramagnets doped with non-magnetic impurities,
which generate a spin polarization cloud as discussed in detail in
Sec.~\ref{sec:1imp}.
This spin polarization can be made static by applying an external field
larger than the interimpurity interaction (see Sec.~\ref{sec:imp-int} below).
The method of choice would be spin-resolved scanning tunneling
microscopy (STM) imaging of the cloud
which, however, is not readily available with atomic resolution.
Alternatively, nuclear magnetic resonance provides a tool to detect the distribution
of local fields.
For instance, for doped antiferromagnetic spin
chains,\cite{takigawa} a broadening of the NMR spectrum at very low
temperatures is observed while, for a doped Haldane chain, the
satellite peaks in the NMR spectrum even allow the experimental
determination of the magnetic correlation length.\cite{tedoldi}

The Knight shift $K_i$ of the resonance field at the site $i$ is proportional to the
local electronic moment (omitting the factor of the electron-nucleus hyperfine coupling);
the NMR spectrum simply corresponds to the distribution of the local shifts $K_i$.
We are interested in the staggered component of the polarization,
which can be defined in 1D as\cite{poilblanc03}
\begin{equation}
\langle S^Z_i \rangle_{\rm stag} = \frac{1}{4}(-1)^i\left[
          2\langle S^Z_{i} \rangle - \langle S^Z_{i+1} \rangle
         -  \langle S^Z_{i-1} \rangle  \right].
\label{stagg}
\end{equation}
For the spin chain of Sec.~\ref{sec:dmrg2}, we calculate the distribution of staggered
Knight shifts using
\begin{equation}
I(x) = \sum_{i=1}^{2N} f_\epsilon\left[x - \langle S^Z_i
  \rangle_{\rm stag}\right],
\label{nmr}
\end{equation}
where $f_\epsilon(x)$ is chosen as a Lorentzian with a small width
$\epsilon$ in order to have a smooth $I(x)$ curve.

\begin{figure}[t]
\centerline{\includegraphics[clip,width=2.7in]{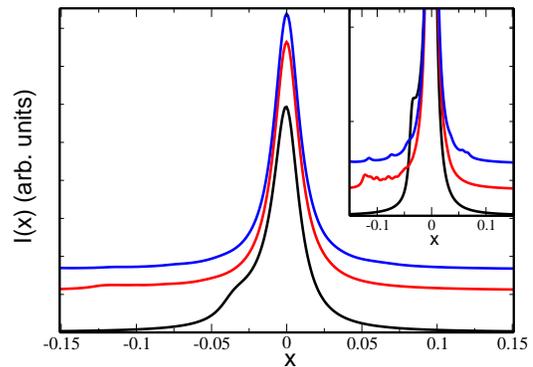}}
\caption{(color online)
   Distribution function $I(x)$ (NMR spectrum) with broadening $\epsilon = 0.02$
   for $J_2=0.48$ and several values of $\delta$ calculated using the
   DMRG results of Sec.~\ref{sec:dmrg2}. The curves are vertically
   shifted for better visualization. From bottom to top we have
   $\delta = 0.0001$ (black), $0.004$ (red), and $0.04$ (blue).
   Inset: $I(x)$ with a smaller broadening $\epsilon = 0.01$.
}
\label{nmr-spec}
\end{figure}

In Fig.~\ref{nmr-spec} we show examples of the NMR spectrum calculated
using Eq.~\eqref{nmr} and the DMRG results derived in
Sec.~\ref{sec:dmrg2} for a chain with a single impurity.
The results are for $J_2 = 0.48$ and $\delta =
0.0001$, $0.004$, and $0.04$ which are in the regions of weak,
intermediate, and strong confinement, respectively. As it was already
observed in Ref.~\onlinecite{elbio} the main effect is the
broadening of the spectral line.

Weak confinement leads to an asymmetric spectrum, with a pronounced shoulder for one sign
of $x$. This asymmetry is directly related to the spatial asymmetry of the spin
polarization cloud and can be understood as follows: for an impurity
as in Fig.~\ref{imp-ham2}, sites with $i > \imp$ ($i<\imp$) have
$\langle S^Z_i \rangle_{\rm stag} < 0$ ($\langle S^Z_i \rangle_{\rm stag} > 0$), respectively.
Hence, the part of the cloud dominated by weak confinement, $i > \imp$,
contributes to $I(x<0)$, where the large extension of the cloud results in
a prominent tail of $I(x<0)$.
Obviously, for a finite concentration of impurities, evenly distributed on both
sublattices, the asymmetry of the line shape will disappear, but the shoulder feature
remains.


\subsection{Effective interaction between vacancy-induced moments}
\label{sec:imp-int}

In the experimentally relevant situation of a finite concentration of non-magnetic
impurities, the induced moments couple via an indirect exchange interaction mediated by
the gapped host. If the underlying lattice is bipartite, then the arising magnetism
is typically nonfrustrated, despite the strong disorder inherent to the problem.
In $D=3$ and at sufficiently low temperatures, this interaction may lead to
impurity-induced magnetic order in the paramagnetic host.
Experimentally, this phenomenon has been observed in a variety of magnets, e.g.,
Zn-doped CuGeO$_3$,\cite{hase96,martin97,regnault95} Zn-doped SrCu$_2$O$_3$,\cite{azuma97}
and Mg-doped TlCuCl$_3$.\cite{sitetlcucl}

On the theoretical side, vacancy-induced magnetic ordering in such compounds is commonly
described by effective models, which exclusively contain the impurity-induced moments
of spin 1/2, which are the only low-energy degrees of freedom below the bulk spin gap
(for small impurity concentration).
The coupling between these moments is written down as a Heisenberg-type model,
with spins $S_i$ at the random impurity locations, and a pairwise
exchange couplings $J_{ij}$ of the form
\begin{equation}
J_{ij} \propto (-1)^{r_{ij}}\exp(-r_{ij}/\lmag).
\label{jeffetivo}
\end{equation}
Here $r_{ij}$ is the distance between two impurities at sites $i$ and
$j$. Note that the common assumption is that the decay length of the $J_{ij}$ is the bulk
correlation length $\lmag$.

In the following, we argue that Eq.~\eqref{jeffetivo} does not apply to the regime of weak
confinement: there, the decay length of the $J_{ij}$ is given by $\lconf$ instead.
The simplest argument is that $J_{ij}$ is determined by the overlap
between the spin polarization clouds associated with the vacancies at
sites $i$ and $j$. As shown above, the cloud size crosses over from $\lmag$ in the
strongly confined regime to $\lconf$ in the weakly confined regime -- the same now
applies to the decay length of the $J_{ij}$.

\begin{figure}[t]
\centerline{\includegraphics[clip,width=3.4in]{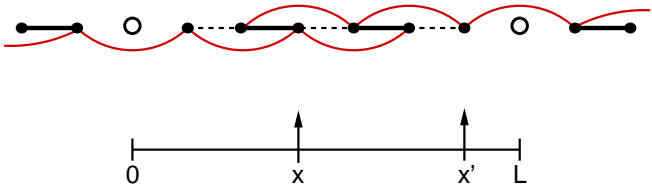}}
\caption{(color online)
Schematic representation of the Hamiltonian
   \eqref{hchain} with two non-magnetic impurities (top panel) and two
   spinons between two impurities separated by a distance $L$
   (bottom panel). The symbols are the same as in Fig.~\ref{ham-spinons}.
}
\label{imp-imp}
\end{figure}

To make this explicit, let us consider the spin chain described by the
Hamiltonian \eqref{hchain} with two vacancies, the first one located
at an odd site and the second at an even site, as illustrated in
Fig.~\ref{imp-imp}.
In this configuration, the two spinons stay mainly in the
region of size $L$ between the two impurities.
As for the strongly confined regime, we assume that
in the ground state the two spinons are in a singlet configuration
with energy $E^0_{S=0}$ and define the effective exchange coupling as
\begin{equation}
J^{\rm eff} = E^0_{S=1} - E^0_{S=0},
\label{jefetivo2}
\end{equation}
where $E^0_{S=1}$ is the energy of
the lowest triplet configuration.\cite{poilblanc03}

We proceed along the lines of Refs.~\onlinecite{uhrig,byrnes} and study the
two-impurity problem near the disorder line
via an effective Schr\"odinger equation in the continuum limit,
which captures the physics in the Hilbert space sector without excited triplets.
The continuum limit is justified here, as the effective interaction
turns out to be dominated by the long-distance tail of the wavefunction.
Within this approach, the difference between
the singlet and the triplet configurations is given by the boundary
conditions imposed on the total wavefunction $\Psi(x,x')$. For the
triplet configuration, the two spinons are not allowed to penetrate
through each other while such process is permitted for the singlet
one. Moreover, if we assume that the total wavefunction can be written
as a product, i.e., $\Psi(x,x') \sim \psi(x)\psi(x')$, then we can
reduce the two-particle problem to a single-particle one. Such
considerations lead us to describe a spinon of mass $m$ at position
$x$ by the Schr\"odinger equation
\begin{equation}
E\psi(x) = -\frac{J}{2m}\frac{\partial^2}{\partial x^2}\psi(x) +
           \frac{3\delta J}{4}x\psi(x),
\label{sch-eq}
\end{equation}
subject to the following boundary conditions
\begin{eqnarray}
{\rm singlet:} && \psi^s(x=0) = \psi^s(x=L) = 0,
\label{bound-sing}\\
&& \nonumber \\
{\rm triplet:} && \psi^t(x=0) = \psi^t(x=L/2) = 0.
\label{bound-trip}
\end{eqnarray}
Notice that Eq.~\eqref{bound-sing} reflects the fact that the motion
of the spinon is restricted to the region between the two impurities
while the second term of Eq.~\eqref{bound-trip} simulates the
nonpenetration condition for the triplet configuration.

\begin{figure}[t]
\centerline{\includegraphics[clip,width=2.7in]{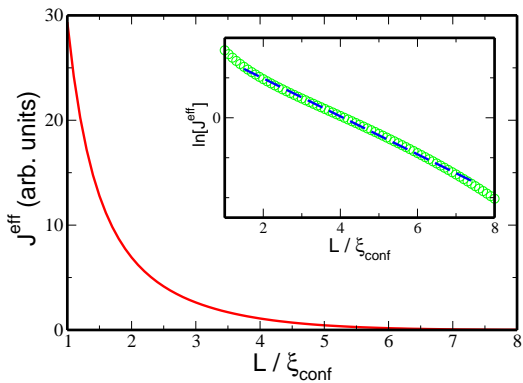}}
\caption{(color online)
   Effective exchange coupling $J^{\rm eff}$ [Eq.~\eqref{jefetivo2}] as a
   function of the renormalized impurity--impurity distance
   $L/\lconf$. Inset: Natural logarithm of $J^{\rm eff}$
   (open green circles) and the corresponding
   linear fitting (dashed blue line).
 }
\label{fig:jefetivo}
\end{figure}

The differential equation \eqref{sch-eq} has an analytical solution given in terms of a
linear combination of shifted Airy functions ${\rm Ai}(x)$ and ${\rm
Bi}(x)$, i.e.,\cite{handbook}
\begin{equation}
\psi(x) = \alpha{\rm Ai}(x/\xi + x_0) + \beta{\rm Bi}(x/\xi + x_0),
\label{solution-sch}
\end{equation}
with the corresponding eigenvalue
\begin{equation}
E = -\frac{Jx_0}{2m\xi^2}. 
\end{equation}
Based on the discussions on Sec.~\ref{sec:var},
$\xi = (3m\delta/2)^{-1/3}$ is identified with $\lconf$.
$x_0$ is determined from the boundary conditions \eqref{bound-sing}
and \eqref{bound-trip}, more precisely, from the solutions of the
following equation
\begin{equation}
{\rm Bi}(x_0){\rm Ai}(l+x_0) - {\rm Ai}(x_0){\rm Bi}(l+x_0) = 0,
\label{root}
\end{equation}
with $l=L/\lconf$ (singlet) and $l=L/(2\lconf)$ (triplet). We solve
Eq.~\eqref{root} numerically, calculate the corresponding effective
exchange coupling as a function of $L/\lconf$ according to
Eq.~\eqref{jefetivo2}, and plot the result in
Fig.~\eqref{fig:jefetivo}. One can see that $J^{\rm eff}$ decays
exponentially with the renormalized impurity--impurity distance $L/\lconf$.
Indeed, the log-linear data in the inset of Fig.~\eqref{fig:jefetivo}
can be fitted by $\ln(J^{\rm eff}) = a_0 - a1(L/\lconf)$ with $a_0 = 3.8681$
and $a_1 = 0.9491$.

As the calculation applies to a parameter regime of the spin chain where the bulk
correlation length is of order unity, this shows that the effective interaction
follows Eq.~\eqref{jeffetivo} with the replacement $\lmag\to\lconf$.

\subsection{Vacancy-induced order: Staggered magnetization}

In a situation where the effective interaction discussed above leads
to long-range magnetic order, the staggered magnetization, being the
order parameter, will be strongly inhomogeneous. Nevertheless, the
Bragg peak seen in the elastic neutron scattering signal at the
ordering wave vector $\vec Q$ will be proportional to the {\it total}
staggered magnetization of the system. For dilute impurities and
neglecting fluctuation effects, this is just given by the number of
impurities times the staggered magnetization per impurity, $M_{\rm
stag}$. Note that this quantity can be much larger than 1/2, in
contrast to the uniform moment associated with each impurity which
equals 1/2.

\begin{figure}[t]
\centerline{\includegraphics[clip,height=5.7cm]{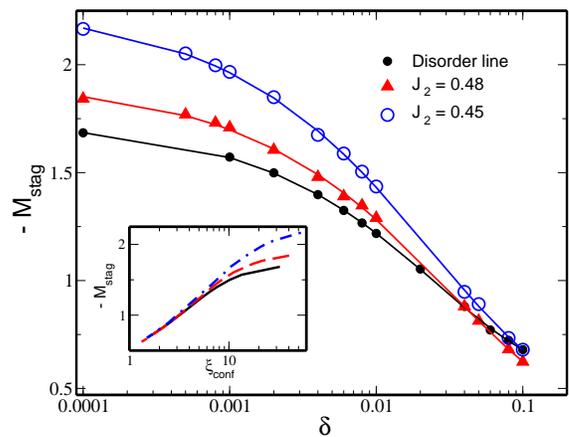}}
\caption{(color online)
The total staggered magnetization associated with
   one impurity as a function of $\delta$ for the disorder, $J_2 =
   0.48$ and $0.45$ lines calculated
   using the DMRG results of Sec.~\ref{sec:dmrg2}. The solid lines are
   fitted to the data (see text for details). Inset: ${\rm M_{\rm stag}}$
   in terms of $\lconf$. Results for the
   disorder (solid black line), $J_2 =
   0.48$ (dashed red line), and $J_2 = 0.45$ (dashed-dot blue line) lines.
}
\label{staggered-mag}
\end{figure}

In the following, we calculate the staggered magnetization
\begin{equation}
M_{\rm stag} = \sum_{i=1}^{2N}(-1)^i\langle S^Z_{i} \rangle,
\label{stag}
\end{equation}
for a spin chain with a single vacancy from the DMRG data of
Sec.~\ref{sec:dmrg2}. The behavior of $M_{\rm stag}$ as
a function of $\delta$ for the disorder, $J_2 = 0.48$ and $0.45$ lines
is displayed in Fig.~\ref{staggered-mag}. One can see that $|
M_{\rm stag}|$ increases as we move into the weakly confined regime
but does {\em not} diverge in the limit $\delta \rightarrow 0$,
despite the diverging cloud size, $\lconf\to\infty$, in this limit.
Indeed, by fitting the curves with the expression
\[  -M_{\rm stag}(\delta) = c_0 + c_1\exp(-c_2\delta^{c_3}), \]
we find that $M_{\rm stag}(\delta \rightarrow 0) =  1.7275,$
$1.9186$, and $2.2648$, respectively for the disorder, $J_2 = 0.48$,
and $0.45$ lines.

The finite value of $M_{\rm stag}$ as $\lconf\to\infty$ is to be contrasted
with the situation of an impurity near a magnetic quantum critical point in $2\leq D\leq4$,
where $\lmag\to\infty$ leads to $M_{\rm stag}\to\infty$.\cite{vbs}
Again, the difference can be understood within the picture developed above:
on the one hand, for $\lconf\to\infty$ and finite $\lmag$, the polarization cloud is
primarily made out of many magnetically neutral singlets, see Fig.~\ref{imp-ham2}.
Indeed, in this case, it is possible to show
analytically\cite{uhrig} that $M_{\rm stag}$ approaches a constant
as $\delta\to 0$.
(This constant, however, depends on $\lmag$, see Fig.~\ref{staggered-mag}.)
On the other hand, for $\lmag\to\infty$, the cloud consists of abundant
triplets, which lead to $M_{\rm stag}\to\infty$.

\section{Higher dimensions}
\label{sec:2d}

So far, we have discussed one-dimensional spin systems where
deconfinement occurs generically in models with spin 1/2 per unit cell,
and confinement has to be explicitly induced by dimerization.
The situation in $D\geq2$ is quite different, as deconfinement is not generic.
Instead, it has been shown that the low-energy physics of many spin models
can be written in terms of gauge-field theories.
Here, the constituents are fractionalized particles (spinons)
being minimally coupled to a gauge field of Z$_2$ or U(1)
symmetry.\cite{lee,ss04,ss08}
The gauge theory may have both deconfining and confining phases.
While the former display true fractionalization,
the gauge-field dynamics in the latter leads to confinement
of objects carrying a gauge charge (e.g. spinons).
The deconfined phases display topological order,
characterized by the suppression of topological defects of the
gauge field and by non-trivial ground state degeneracies.
In contrast to 1D systems, non-trivial quantum phase transitions
between confined and deconfined phases can be driven by varying microscopic model
parameters.

This discussion makes clear that the physical mechanisms for deconfinement in 1D and
higher dimensions are quite
different. However, in both cases, a system can be in a
weakly confined regime, i.e., close to deconfinement, with qualitatively similar
observable consequences. In the following, we briefly sketch which of our results can be
directly carried over to $D\geq 2$.

\subsection{Confined and deconfined paramagnets}

In spin models with spin 1/2 per unit cell, quantum paramagnets can come in different
varieties: valence-bond solid phases spontaneously break the lattice translation symmetry
and display gapped spin-1 excitations, i.e., confined spinons. In contrast, spin liquids
do not break any symmetry of the Hamiltonian and feature deconfined spin-1/2 spinons,
which can be gapless or gapped. In addition, spin liquids display low-lying gauge-field
excitations: in the Z$_2$ case, these are gapped vortices of the Z$_2$ gauge field,
dubbed ``visons'', whereas the U(1) case displays a linearly dispersing ``photon'' mode.
These excitations carry zero spin, i.e., translate into singlet excitations in terms of
the original spin variables.

Weak confinement can for instance be expected
(i) near a paramagnetic confinement--deconfinement transition, e.g.,
from a VBS to a Z$_2$ spin liquid on the triangular lattice \cite{sr91,xuss09,vbs_z2,qdm},
and (ii) near a deconfined critical point, e.g., a transition from a VBS to a N\'eel state on
the square lattice.
In both cases, the energy scale associated to confinement will be small in the VBS phase
close to the quantum critical point.
Consequently, there will be a large length scale $\lconf$, with $\lconf\gg\lmag$,
which characterizes the binding of spinons, which may be detected using nonmagnetic
impurities as described above.
In a VBS phase, $\lconf$ also plays the role of a characteristic domain wall
thickness.\cite{deconf}

Candidates for a deconfined VBS--N\'eel transition are the $J_1$--$J_2$ and
$J$--$Q$ square lattice spin-1/2 Heisenberg models, where the antiferromagnetic order is
frustrated either by a second-neighbor ($J_2$) or ring-exchange ($Q$) interaction.
Whereas the transition in both cases may be weakly first order, observable finite-temperature
properties of the $J$--$Q$ model have been found to be in agreement with predictions of deconfined
criticality.\cite{kaul}

Indications for weak confinement have indeed been found in a numerical study \cite{poilblanc1}
of the frustrated $J_1$--$J_2$--$J_3$ Heisenberg model on the square lattice, where
the couplings $J_i$ connecting $i$-th neighbor sites are all antiferromagnetic.
For $J_3=0$, this model displays a paramagnetic phase for $0.4 < J_2/J_1 < 0.6$,
which has been proposed to be of columnar\cite{read} or plaquette\cite{ueda} VBS type.
For $J_2=0$, the paramagnetic phase appears to be larger,  $0.3 < J_3/J_1 < 0.7$,\cite{leung,capriotti}
and again columnar\cite{leung} and plaquette\cite{mambrini,arlego} VBS states
have been suggested.
Ref.~\onlinecite{poilblanc1} studied the $J_1$--$J_2$--$J_3$ model using exact diagonalization,
near the line $J_2 + J_3 = J_1/2$ where the VBS order is likely most stable.
Interestingly, these calculations showed clear indications of the existence of
two length scales: while the magnetic bulk correlation length, $\lmag$, was
rather short, a significantly larger scale could be identified in the decay of the spin
polarization near a vacancy -- consistent with our analysis, this is to be identified
with the confinement length $\lconf$.
For instance, for $J_{1,2,3} = (1,0.3,0.2)$, $\lmag\approx0.7$ and $\lconf\approx5.5$.\cite{poilblanc1}
These findings
(i)  suggest that weak confinement is common to many frustrated low-dimensional magnets, and
(ii) support that the paramagnetic phase of the $J_1$--$J_2$--$J_3$ model is located near
a deconfined critical point.

Among the various differences between 1D and higher dimensions is that
weak confinement in $D\geq 2$ implies low-lying singlet excitations,
which are energetically located below the triplet gap;
in the gauge-field language, these are, e.g., visons near a confinement--deconfinement
transition.
In contrast, in the $J_1$--$J_2$--$\delta$ chain there are no low-lying singlets, as any
excitation requires the creation of two spinons with an energy cost of at least the
triplet gap.


\subsection{Implications of linear confinement}

In the language of spinons coupled to gauge fields, a treatment of the weakly confined
regime is difficult, as the gauge-field effects leading to confinement are
non-perturbative. In the following, we simply {\em assume} that the gauge field induces
weak linear pairwise confinement between spinons (as well as between spinons and holons)
and discuss its consequences.
In fact, linear confinement is a plausible result of VBS formation in $D\geq
2$.\cite{deconf}
Note, however, that bound states can in principle exist in a situation of
asymptotic deconfinement, see Sec.~\ref{asympt}.

The simplest possible analysis employs a linear confinement potential $V(\br) = \alpha r$ in the
continuum limit.
Consider a confined paramagnet with a single vacancy, the latter
introducing a spinon into the system. Assuming the spinon to be massive with quadratic
dispersion and to be linearly bound to the vacancy, its motion is described by
a $D$-dimensional version of Eq.~\eqref{sch-eq}:
\begin{equation}
     -\frac{J}{2m}{\mathbf \nabla}^2\Psi(\br)
              + \alpha J r\Psi(\br) = E\Psi(\br).
\label{sch-eq2}
\end{equation}
Here $r$ is the radial distance from $\imp$.
The solution of the above differential equation can be written as
\begin{equation}
\Psi(\br) = \left\{ \begin{array}{lr}
               \psi(r)\exp(-il\theta), & D=2, \\
               \psi(r)Y_l^m(\theta,\phi), & D=3,
                   \end{array}\right.
\label{2d-sol}
\end{equation}
where $Y_l^m(\theta,\phi)$ is a spherical harmonic.\cite{handbook}
Unfortunately, the resulting radial problem has no exact
(analytical) solution. We then proceed, assuming that the wavefunction
of the lowest energy vacancy--spinon bound state can be approximated
by
\begin{equation}
\psi(r) = (r/\xi)^\beta\exp\left(-(r/\xi)^\beta\right),
\label{wf2d}
\end{equation}
where $\xi = (2m\alpha)^{-1/3}$ and the coefficient $\beta$ is
variationally determined. This choice is again motivated by the
asymptotic form of the Airy function ${\rm Ai}(x)$ [see
Eq.~\eqref{fit}]. It also satisfies the boundary conditions
$\psi(0)=0$ and $\psi(\infty)=0$. By minimizing the total energy $E =
\int d\br \Psi^*(\br)H\Psi(\br)/\int d\br |\Psi(\br)|^2$, with $H$
given by Eq.~\eqref{sch-eq2}, it is easy to show that, for $D=2$ and
$l=0$, we have
\[
   E = \frac{J}{2m\xi^2}\frac{\beta^2/8 + 2^{-2 -3/\beta}\Gamma(2 + 3/\beta)}
                  {2^{-2(1+\beta)/\beta}\Gamma(2 + 2/\beta)},
\]
where $\Gamma(x)$ is the Gamma function.\cite{handbook} The above
expression assumes a minimum value $E = 2.1967\,J/(2m\xi^2)$ for
$\beta = 1.1989$. Similarly, for $D = 3$, one finds $E =
2.5187\,J/(2m\xi^2)$ and $\beta = 1.2682$.
It is worth mentioning that the same variational procedure applied
to the one-dimensional case gives $\beta = 1.1428$
and $E = 2.4429\,J/(2m\xi^2)$, which is a good approximation for the exact
analytical solution of Eq.~\eqref{sch-eq2}, $E = 2.3381\,J/(2m\xi^2)$,
calculated in Ref.~\onlinecite{uhrig}.

We conclude that the vacancy physics in the weakly confined regime in $D\geq 2$
is also dominated by the confinement length: Eq.~\eqref{wf2d} has a maximum
for $r=\xi$, which can be identify with $\lconf$. As we move deep into
the weakly confined region (smaller $\alpha$), $\xi$ increases. The spin
polarization cloud around $\imp$, whose characteristic length scale is
$\xi = \lconf$, does not display a pure exponential decay, but instead
has a profile similar to the 1D case. This also suggests that
analyzing the numerical data in terms of an exponential decay, as done
in Ref.~\onlinecite{poilblanc1}, might be not appropriate in the
weakly confined regime.

Two-spinon bound states can be addressed similarly.
The starting point is the $D$-dimensional version of Eq.~(10)
from Ref.~\onlinecite{byrnes}, which is nothing but
Eq.~\eqref{sch-eq2} with the replacements: $m \rightarrow \mu = m/2$
and $E \rightarrow E - \Delta$. Here $\mu$ is the reduced mass,
$\Delta$ is the spin gap (the energy of the lowest bound
state), and $r$ is the relative radial distance between the two
spinons. As a consequence, the characteristic length scale is now $\xi
= (2\mu\alpha)^{-1/3}$.
The solutions of the effective Schr\"odinger equation, which describes
the two-spinon bound states {\it above} the spin-gap $\Delta$, also have the
form \eqref{2d-sol}. We assume that Eq.~\eqref{wf2d} is a good
approximation for the radial part of the first bound-state
wavefunction $\psi_1(r)$ and variationally
determine the parameter $\beta$ and the energy $E_1$ for the case
$l=0$. Once $\beta$ is known,
variational wavefunctions for higher bound states can be easily
constructed just by multiplying $\psi_1(r)$ by a polynomial, e.g.,
\begin{eqnarray}
\psi_2(r) &=& [(r/\xi)^2 + c_{21}]\psi_1(r), \nonumber \\
\psi_3(r) &=& [(r/\xi)^4 + c_{32}(r/\xi)^2 + c_{31}]\psi_1(r),
\label{wf2d-exc}
\end{eqnarray}
respectively, for the second and third lowest-energy bound state above
$\Delta$. The coefficients $c_{ij}$ are determined by requiring that
the states are orthogonal to each other. Notice that
Eq.~\eqref{wf2d-exc} satisfy the boundary condition $\psi(0) = 0$ and
they vanish for large $r$. The {\it ansatz} \eqref{wf2d-exc} follows
from a careful analysis of the analytical solutions of the
one-dimensional problem. Indeed, we applied the above procedure for the
$D=1$ case, compared the results with the analytical
solution reported in Ref.~\onlinecite{byrnes} and verified that it
provides good estimates for the energy of the lowest two-spinon
bound states.

For $D=2$ and $l=0$, we find again $\beta = 1.1989$ and that the
energies of the first ($E_1$), second ($E_2$) and third ($E_3$) lowest
bound states are given by $E_i = e_i + \Delta$ with $e_1 = 2.1967$,
$e_2 = 4.0025$, and $e_3 = 5.4875\,J/(2\mu\xi^2)$.
According to the discussion in Sec.~\ref{dynamics1}, the
strong--weak confinement crossover occurs when $e_1 \sim \Delta$.
Similar considerations hold for $D=3$.

The above analysis suggests many parallels between 1D and higher dimensions,
but falls short of capturing short-distance features of the bound states as well as
effects of non-trivial statistics of the spinons.


\subsection{Asymptotic confinement vs. short-range bound states}
\label{asympt}

The preceding discussion of confinement has been simplistic in that it assumed that the
presence or absence of long-distance (asymptotic) confinement also determines the nature
of the low-energy excitations. While this is the case for the one-dimensional
$J_1$--$J_2$--$\delta$ spin chain, which features a confinement potential of the
type $V(x) = \alpha|x|$ for all length scales, this is obviously not true in general.
While the long-distance behavior of the effective interaction, mediated by the gauge
field, decides about asymptotic confinement or deconfinement, it is its short-distance
behavior which determines the existence and spatial size of possible bound states at
low energies.

In other words, it is conceivable that a phase in $D\geq 2$ spatial
dimensions with asymptotic deconfinement of spinons and associated
topological order displays short-distance spinon bound
states.\cite{ed_foot} In such a situation, the diagnostics proposed in
this paper, i.e., an vacancy-induced magnetic moment with cloud size
$\lconf$ and multiple sharp modes in $\chi''$ below the continuum,
will signal (weak) confinement, whereas the true nature of the phase
is deconfined. Obviously, distinguishing such a deconfined phase with
bound states from a confined phase is extremely difficult.
(Numerically, topological ground-state degeneracies may be used to
detect deconfinement.)

To our knowledge, no explicit microscopic examples for fractionalized spin liquids with
spinon bound states existing below the continuum are known to date,\cite{vidal09}
which is also related to the difficulties in numerically studying the excitation spectra
of the relevant microscopic models.
Possible candidates are states with $Z_2$ fractionalization
which are believed to generically display a non-zero spinon pairing amplitude.\cite{sf}
It should be kept in mind that the bound states in question, which are determined by short-range
physics, are not necessarily part of effective low-energy gauge-field descriptions.

It is worth noting that a spinless impurity in the VBS phase near a deconfined critical
point has been shown\cite{kolezhuk} to bind a spinon on a length scale much shorter
than the VBS (or confinement) length.


\section{Summary}
\label{conclusion}

In this paper, we have argued that quantum paramagnets, in particular those with
frustration, can display an interesting regime of weak confinement of spinons.
Here, the long-distance properties are conventional, with elementary excitations being
spin-1 triplons; however, the physics of spinons is visible at shorter length scales.
Such weakly confined paramagnets are characterized by two potentially large length
scales, namely the common magnetic correlation length $\lmag$ and a confinement length
$\lconf$, which measures the typical distance between weakly bound spinons. Weak
confinement implies $\lconf\gg\lmag$; the crossover to strong confinement occurs at
$\lconf\approx\lmag$.

We have illustrated the physics of the weakly confined regime with explicit calculations
for frustrated dimerized spin chains, where weak confinement can be realized for weak explicit
dimerization on top of strong spontaneous dimerization.
In particular, we have studied the static response of the system to
nonmagnetic impurities: the staggered
polarization cloud induced by the impurity decays on a length scale given by ${\rm max}(\lconf,\lmag)$.
We have also briefly discussed the structure of the dynamic susceptibility in the clean case
and the effective interaction between impurity-induced moments.
Qualitatively, our results remain valid in higher dimensions as well:
in $D\geq 2$, weak confinement is reached near a confinement--deconfinement transition,
e.g., between a gapped spin liquid and a valence-bond solid, or near a deconfined quantum
critical point between a N\'eel and a VBS phase.

However, a caveat in $D\geq 2$ is that long-distance confinement (deconfinement)
of spinons is not necessarily equivalent to the presence (absence) of short-distance spinon bound
states: In principle, bound states forming conventional triplon excitations can exist
even in deconfined spin liquids. This tantalizing scenario is left for future
investigations.

Last not least, upon inclusion of mobile charge carriers, weak confinement
can be expected to lead to an almost spin--charge separated metallic state.
This may be a conventional Fermi liquid, but with small quasiparticle weights and
strong incoherent continua in spectral functions:
The carriers will get dressed by a large cloud of {\em singlet} excitations,
in analogy to the dressing with {\em triplet} excitations near a magnetic quantum
critical point.\cite{soc}


\acknowledgments

We thank R. Moessner, A. Rosch, S. Sachdev, U. Schollw\"ock,
T. Senthil, and G. Japaridze for discussions and S. Trebst for
suggestions. This research was supported by the DFG through
SFB 608 (K\"oln).


\appendix

\section{Variational approach for the disorder line}
\label{ap:variational}

In this section we present some details of the variational calculations
discussed in Sec.~\ref{sec:var}.

We start by looking at the overlap between the states $|m\rangle$ and
$|p\rangle$. By comparing the two states one can see that there is a
mismatch in the position of the singlets in the region between the
sites $2m$ and $2p$. Each mismatch contributes with a factor $(-1/2)$
and the number of mismatches is equal to $|m-n|$. Thus,
Eq.~\eqref{overlap} follows.

In order to calculate the matrix elements \eqref{matrix} it is
convenient to introduce the operator\cite{majumdar}
\begin{equation}
h(i,j) = \frac{1}{2}\left(1 - 4{\mathbf S}_i\cdot{\mathbf S}_j
\right).
\label{oph}
\end{equation}
By applying the operator $h(i,j)$ to a normalized singlet combination
$[i,j]$ or a product of them, we have
\begin{eqnarray}
h(i,j)[i,j] &=& 2[i,j],
\nonumber \\
h(i,j)[k,i][j,l] &=& (-1)[k,l][i,j],
\label{prop-h} \\
h(i,j)[k,i][l,j] &=& [k,j][i,l] + [k,i][l,j].
\nonumber
\end{eqnarray}
In the presence of an
unpaired spin $|\uparrow\rangle_i$ the following expressions hold
\begin{eqnarray}
h(i,j)[l,i]|\uparrow\rangle_j &=& (-1)|\uparrow\rangle_l[i,j],
\nonumber \\
h(i,j)|\uparrow\rangle_i[j,l] &=& (-1)[i,j]|\uparrow\rangle_l,
\label{prop-h2} \\
h(i,j)[i,l]|\uparrow\rangle_j &=& |\uparrow\rangle_i[l,j] +
                                  [i,l]|\uparrow\rangle_j.
\nonumber
\end{eqnarray}

Instead of the Hamiltonian \eqref{hchain} with a nonmagnetic impurity
at $\imp = 1$, we consider an auxiliary one $H'$ which is defined as
\begin{equation}
H' = -2H + (J_1 + J_2)(N - 1).
\label{hprime}
\end{equation}
The above equation can be written in terms
of the operator $h(i,j)$ \eqref{oph}, namely ($J_1=1$)
\begin{eqnarray}
H' &=& (1+\delta)\sum_{i=2}^N h(2i-1,2i)
    + (1-\delta)\sum_{i=1}^{N-1}h(2i,2i+1)
\nonumber \\
   &+&   J_2\sum_{i=2}^{N-1} h(2i-1,2i+1) + J_2\sum_{i=1}^{N} h(2i,2i+2).
\nonumber \\
\label{hprime2}
\end{eqnarray}

Using the properties \eqref{prop-h} and \eqref{prop-h2} we can
calculate $H'|p\rangle$. For instance, the contribution of the first term of
Eq.~\eqref{hprime2} reads
\begin{equation}
\sum_{i=2}^N h(2i-1,2i)|p\rangle = 2(N-p)|p\rangle
            - |p - 1\rangle
            - \sum_{i=2}^{p-1}|i,p\rangle,
\nonumber
\end{equation}
The state $|i,p\rangle$ is quite similar to $|p\rangle$ apart from the
fact that a long singlet is introduced in the vicinity of the site
$2i$, i.e.,
\begin{eqnarray}
|i,p\rangle &=& [1,2][3,4]\ldots[2i-1,2i][2i-2,2i+1]\ldots
\nonumber \\
            &\ldots& |\uparrow\rangle_{2p}\ldots[2N-3,2N-2][2N-1,2N].
\nonumber
\end{eqnarray}
For the second term of Eq.~\eqref{hprime2} we have
\begin{equation}
\sum_{i=1}^{N-1} h(2i,2i+1)|p\rangle = 2(p-1)|p\rangle
               - |p + 1\rangle
               - \sum_{i=p+1}^{N-1}|i,p\rangle'
\nonumber
\end{equation}
with
\begin{eqnarray}
|i,p\rangle' &=& [1,2][3,4]\ldots|\uparrow\rangle_{2p}\ldots
\nonumber \\
            &\ldots& [2i,2i+1][2i-1,2i+2]\ldots[2N-1,2N].
\nonumber
\end{eqnarray}
Notice that $|i,p\rangle$ and $|i,p\rangle'$ are similar, but for the
former we have $i < p$ while for the latter, $i > p$. The results for
the two remaining terms can also be expressed in terms of $|p\rangle$,
$|p\pm 1\rangle$, $|i,p\rangle$ and $|i,p\rangle'$.
With the help of \eqref{hprime} it is possible to show that
\begin{eqnarray}
  && 2 (H - \bar{E}_0)| p \rangle = (4p\delta + 1 + 3J_2 - 2\delta)| p \rangle
     - J_2|\chi_p\rangle
\nonumber \\
\nonumber && \\
  &&+ (1 + \delta - J_2)| p - 1\rangle
    + (1 - \delta - J_2)| p + 1\rangle \nonumber \\
\nonumber && \\
  &&+ (1 + \delta - 2J_2)\sum_{i=1}^{p-1}|i,p\rangle
    + (1 - \delta - 2J_2)\sum_{i=p+1}^{N-1}|i,p\rangle'
\nonumber \\
\label{aux-matrix}
\end{eqnarray}
where $\bar{E}_0 = -(1/4)(1 + 2\delta + J_2)(2N)$ and the state
$|\chi_p\rangle$ is given by
\begin{eqnarray}
|\chi_p\rangle &=& (-1)[1,2][3,4]\ldots[2p-2,2p+2][2p-1,2p+1]
\nonumber \\
&& \nonumber\\
            && |\uparrow\rangle_{2p}\ldots[2N-3,2N-2][2N-1,2N].
\nonumber
\end{eqnarray}

Proceeding in the same way as for Eq.~\eqref{overlap}, one can show that
the overlaps between
$|m\rangle$ and the states $|\chi_p\rangle$, $|i,p\rangle$,
and $|i,p\rangle'$ are given by
\begin{eqnarray}
 \langle m |\chi_p\rangle &=& (1/2) \langle m | p \rangle,
\nonumber \\
&& \nonumber \\
  \langle m |i,p\rangle &=& -\frac{1}{2}\langle m
                             |p\rangle\left[\theta(m-p) + \theta(p-m-1)\right.
\nonumber \\
&& \nonumber \\
        && \left. \times\left(\theta(m-i) +
            4\theta(i-m-1)\right)\right],
\nonumber \\
&& \nonumber \\
  \langle m |i,p\rangle' &=& -\frac{1}{2}\langle m
                             |p\rangle\left[\theta(p-m) + \theta(m-p-1)\right.
\nonumber \\
&& \nonumber \\
        && \left. \times\left(\theta(i - m) +
            4\theta(m-i-1)\right)\right],
\label{overlap1}
\end{eqnarray}
where the function $\theta(x)$ is defined as
\begin{equation}
\theta(x) =  \left\{\begin{array}{r@{\quad}l}
             1 & x \ge 0, \\
             0 & x < 0.
            \end{array} \right.
\label{theta}
\end{equation}
Equation \eqref{matrix} follows from Eqs.~\eqref{aux-matrix} --
\eqref{overlap1}.



\begin{thebibliography}{}

\bibitem{giamarchi}
T. Giamarchi, {\sl Quantum Physics in One Dimension},
Clarendon Press, Oxford, 2003.

\bibitem{lee}
P. A. Lee, N. Nagaosa, and X.-G. Wen,
Rev. Mod. Phys. {\bf 78}, 17 (2006).

\bibitem{heinonen}
O.\ Heinonen, ed., {\sl Composite Fermions},
World Scientific, Singapore (1998).

\bibitem{ss04}
S. Sachdev, in: {\sl Quantum magnetism},
Lecture Notes in Physics Vol. 645,
edited by U. Schollw\"ock, J. Richter, D. J. J. Farnell, and R. A. Bishop
(Springer, Berlin 2004).

\bibitem{ss08}
S. Sachdev, Nature Physics {\bf 4}, 173 (2008).

\bibitem{moess01}
R. Moessner and S. L. Sondhi,
Phys. Rev. Lett. {\bf 86}, 1881 (2001).

\bibitem{kitaev03}
A. Kitaev,
Ann. Phys. (N.Y.) {\bf 303}, 2 (2003).

\bibitem{kitaev06}
A. Kitaev,
Ann. Phys. (N.Y.) {\bf 321}, 2 (2006).

\bibitem{triang_ring}
G. Misguich, C. Lhuillier, B. Bernu, and C. Waldtmann,
Phys. Rev. B {\bf 60}, 1064 (1999).

\bibitem{deconf}
T. Senthil, L. Balents, S. Sachdev, A. Vishwanath, and M. P. A. Fisher,
Phys. Rev. B {\bf 70}, 144407 (2004).

\bibitem{sandvik}
A. W. Sandvik, Phys. Rev. Lett. {\bf 98}, 227202 (2007).

\bibitem{poilblanc1}
D. Poilblanc, A. L\"auchli, M. Mambrini, and F. Mila,
Phys. Rev. B {\bf 73}, 100403(R) (2006);
D. Poilblanc, M. Mambrini, A. L\"auchli, and F. Mila,
J. Phys.: Conds. Matter {\bf 19}, 145205 (2007).

\bibitem{chitra}
R. Chitra, S. Pati, H. R. Krishnamurthy, D. Sen, and S. Ramasesha,
Phys. Rev. B {\bf 52}, 6581 (1995).

\bibitem{swapan}
S. Pati, R. Chitra, D. Sen, S. Ramasesha, and H. R. Krishnamurthy,
J. Phys.: Cond. Matter {\bf 9}, 219 (1997).

\bibitem{okamoto}
K. Okamoto and K. Nomura, Phys. Lett. A {\bf 169}, 433 (1992).


\bibitem{affleck97} 
I. Affleck, in {\it Dynamical Properties of Unconventional
Magnetic Systems}, Vol. 349 of {\it NATO Advanced Study
Institute, Series E: Applied Sciences},
edited by A. Skjeltrop and D. Sherrington (Kluwer
Academic, Dordrecht, 1998), pp. 123-131. See also
arXiv:cond-mat/9705127v1.

\bibitem{majumdar}
C. K. Majumdar and D. K. Ghosh, J. Math. Phys. {\bf 10}, 1388 (1969).

\bibitem{shastry}
B. S. Shastry and B. Sutherland,
Phys. Rev. Lett. {\bf 47}, 964 (1981).

\bibitem{sorensen}
E. S. S\o rensen, I. Affleck, D. Augier, and D. Poilblanc,
Phys. Rev. B {\bf 58}, 14 701 (1998).

\bibitem{uhrig}
G. S. Uhrig, F. S. Sch\"onfeld, M. Laukamp, and E. Dagotto,
Eur. Phys. J. B {\bf 7}, 67 (1999).

\bibitem{BSfoot}
Two-spinon bound states can be of $S=0$ or $S=1$ character.
The lowest $S=1$ two-spinon bound state is the conventional triplon excitation.
(For the spin chain, the lowest $S=0$ bound state is the ground state itself.)
In the limit of weak confinement, the energetic difference between singlet and triplet
spinon bound states becomes small for higher excited states, as the magnetic interaction
between two distant spinons is small.

\bibitem{dmrg}
S. R. White, Phys. Rev. B {\bf 48}, 10345 (1993);
U. Schollw\"ock, Rev. Mod. Phys. {\bf 77}, 259 (2005).


\bibitem{alps}
F. Alet {\it et al.} (ALPS collaboration), J. Phys. Soc. Jpn. Suppl. {\bf 74}, 30 (2005);
F. Albuquerque {\it et al.} (ALPS collaboration), J. Magn. Magn. Mater. {\bf 310}, 1187 (2007);
see http://alps.comp-phys.org.


\bibitem{sigrist}
M. Sigrist and A. Furusaki,
J. Phys. Soc. Jpn. {\bf 65}, 2385 (1996).

\bibitem{icmp} S. Sachdev and M. Vojta,
{\sl Proceedings of the XIII International Congress on Mathematical Physics, London},
edited by A. Fokas {\em et al.}, (International Press, Boston 2001).

\bibitem{bruce}
B. Normand and F. Mila, Phys. Rev. B {\bf 65}, 104411 (2002).

\bibitem{Miyazaki} T. Miyazaki, M. Troyer, M. Ogata, K. Ueda, and D. Yoshioka,
J. Phys. Soc. Jpn. {\bf 66}, 2580 (1997).

\bibitem{elbio}
G .B. Martins, M. Laukamp, J. Riera, and E. Dagotto, \prl {\bf 78}, 3563 (1997);
M. Laukamp, G. B. Martins, C. Gazza, A. L. Malvezzi, E. Dagotto,
P. M. Hansen, A. C. Lopez, and J. Riera, \prb {\bf 57}, 10755 (1998).

\bibitem{Yasuda} C. Yasuda, S. Todo, M. Matsumoto, and H. Takayama,
Phys. Rev. B {\bf 64}, 092405 (2001), and references therein.

\bibitem{vbs}
M. Vojta, C. Buragohain, and S. Sachdev,
Phys. Rev. B {\bf 61}, 15152 (2000).

\bibitem{ssmv}
S. Sachdev and M. Vojta,
Phys. Rev. B {\bf 68}, 064419 (2003).

\bibitem{handbook} M. Abramowitz and I. A. Stegun, {\sl Handbook of
    Mathematical Functions} (Dover Publications, New York, 1964).

\bibitem{fukuyama}
H. Fukuyama, T. Tanimoto, and M. Saito,
J. Phys. Soc. Jpn. {\bf 65}, 1182 (1996).

\bibitem{poilblanc03}
N. Laflorencie and D. Poilblanc,
Phys. Rev. Lett. {\bf 90}, 157202 (2003).

\bibitem{byrnes}
T. M. R. Byrnes, M. T. Murphy, and O. P. Sushkov,
Phys. Rev. B {\bf 60}, 4057 (1999).

\bibitem{trebst1}
S. Trebst, H. Monien, C. J. Hamer, W. H. Zheng, and R. R. P. Singh,
Phys. Rev. Lett. {\bf 85}, 4373 (2000).

\bibitem{trebst2}
Weihong Zheng, Chris J. Hamer, Rajiv R. P. Singh,
Simon Trebst, and Hartmut Monien,
Phys. Rev. B {\bf 63}, 144411 (2001).

\bibitem{uhrig04}
K. P. Schmidt, C. Knetter, and G. S. Uhrig,
\prb {\bf 69}, 104417 (2004).

\bibitem{takigawa}
M. Takigawa, N. Motoyama, H. Eisaki, and S. Uchida,
Phys. Rev. B {\bf 55}, 14129 (1997).

\bibitem{tedoldi}
F. Tedoldi, R. Santachiara, and M. Horvatic,
Phys. Rev. Lett. {\bf 83}, 412 (1999).


\bibitem{hase96}
M. Hase, K. Uchinokura, R. J. Birgeneau, K. Hirota, and G. Shirane,
J. Phys. Soc. Jpn. {\bf 65}, 1392 (1996).

\bibitem{martin97}
M. C. Martin, M. Hase, K. Hirota, G. Shirane, Y. Sasago, N. Koide, and K. Uchinokura,
Phys. Rev. B {\bf 56}, 3173 (1997).

\bibitem{regnault95}
L. P. Regnault, J. P. Renard, G. Dhalenne, and A. Revcolevschi,
Europhys. Lett. {\bf 32}, 579 (1995).


\bibitem{azuma97}
M. Azuma, Y. Fujishiro, M. Takano, M. Nohara and H. Takagi,
Phys. Rev. B {\bf 55}, R8658 (1997).

\bibitem{sitetlcucl}
A. Oosawa, T. Ono, and H. Tanaka, \prb {\bf 66}, 020405(R) (2002).

\bibitem{sr91}
S. Sachdev and N. Read, Int. J. Mod. Phys. B {\bf 5}, 219 (1991).

\bibitem{xuss09}
C. Xu and S. Sachdev,
Phys. Rev. B {\bf 79}, 064405 (2009).

\bibitem{vbs_z2}
R. A. Jalabert and S. Sachdev, Phys. Rev. B {\bf 44}, 686 (1991);
S. Sachdev and M. Vojta, J. Phys. Soc. Japan {\bf 69}, Suppl. B, 1 (2000).

\bibitem{qdm}
R. Moessner and S. L. Sondhi, Phys. Rev. B {\bf 63}, 224401 (2001).

\bibitem{kaul}
R. G. Melko and R. K. Kaul,
Phys. Rev. Lett. {\bf 100}, 017203 (2008).


\bibitem{read}
N. Read and S. Sachdev,
Phys. Rev. Lett. {\bf 62}, 1694 (1989).

\bibitem{ueda}
M. E. Zhitomirsky and K. Ueda, Phys. Rev. B {\bf 54}, 9007  (1996).

\bibitem{leung}
P. W. Leung and Ngar-wing Lam,
Phys. Rev. B {\bf 53}, 2213 (1996).

\bibitem{capriotti}
L. Capriotti, D. J. Scalapino, and S. R. White,
Phys. Rev. Lett. {\bf 93}, 177004 (2004).

\bibitem{mambrini}
M. Mambrini, A. L\"auchli, D. Poilblanc, and F. Mila,
Phys. Rev. B {\bf 74}, 144422 (2006).

\bibitem{arlego}
M. Arlego and W. Brenig, Phys. Rev. B {\bf 78}, 224415 (2008).


\bibitem{ed_foot}
The existence of short-distance bound states in the presence of long-distance
deconfinement is well known in quantum electrodynamics: The $(-1/r)$ Coulomb potential of
two opposite charges allows for both scattering and bound states (e.g. the hydrogen
atom).

\bibitem{vidal09}
An exception appears to be the toric code model in a transverse field,
where bound states have been seen in the topologically ordered phase:
J. Vidal, R. Thomale, K. P. Schmidt, and S. Dusuel,
preprint arXiv:0902.3547.

\bibitem{sf}
T.~Senthil and M.~P.~A.~Fisher,
Phys. Rev. B {\bf 62}, 7850 (2000);
Phys. Rev. B {\bf 63}, 134521 (2001).

\bibitem{kolezhuk}
A. Kolezhuk, S. Sachdev, R. R. Biswas, and P. Chen,
Phys. Rev. B {\bf 74}, 165114 (2006).

\bibitem{soc}
S. Sachdev, M. Troyer, and M. Vojta,
Phys. Rev. Lett. {\bf 86}, 2617 (2001).

\end{thebibliography}
\end{document}